\documentclass[a4paper,12pt]{article}

\usepackage{amsmath,amsthm,amssymb}
\usepackage{graphicx}
\usepackage{subfigure}
\usepackage{multirow}
\usepackage{epstopdf,float,caption}
\usepackage[dvipsnames]{xcolor}
\usepackage{amsmath}
\usepackage{breqn}
\usepackage{authblk}
\usepackage[pdftex]{hyperref}
\usepackage[export]{adjustbox}
\usepackage[figurename=Fig.]{caption}
\newcommand{\D}{\displaystyle}
\newcommand\scalemath[2]{\scalebox{#1}{\mbox{\ensuremath{\displaystyle #2}}}}
\newtheorem{Thm}{Theorem}[section]

\title{\large {\bf{The role of quarantine and isolation in controlling COVID-19 hospitalization in Oman}}}
\author {{Maryam Al-Yahyai$^a$}, {Fatma Al-Musalhi$^a$}, {Nasser Al-Salti$^b$}, {Ibrahim Elmojtaba$^a$}}
\affil{$^a$Department of Mathematics, College of Science, Sultan Qaboos University, Muscat, Oman.\\ $^b$Department of Applied Mathematics and Science, National University of Science and Technology, Muscat, Oman}
\begin{document}
\date{}
\maketitle \vspace{-0.7cm}
\noindent \textbf{Abstract}\\
In this paper, we build a mathematical model for the dynamics of COVID-19 to assess the impact of placing healthy individuals in quarantine and isolating infected ones on the number of hospitalization and intensive care unit cases. The proposed model is fully analyzed in order to prove the positivity of solutions, to study the local and global stability of the disease-free equilibria and to drive the basic and control reproduction numbers of the model. Oman COVID-19 data is used to calibrate the model and estimate the parameters. In particular, the published data for the year 2020 is used, when two waves of disease hit the country. Moreover, this period of time is chosen when no vaccine had been introduced, but only the non-pharmaceutical intervention (NPI) strategies were the only effective methods to control the spread and, consequently, control the hospitalization cases to avoid pressuring the health system. Based on the estimated parameters, the reproduction number and contribution of different transmission routes are approximated numerically. Sensitivity analysis is performed to identify the significant parameters in spreading the disease. Numerical simulation is carried out to demonstrate the effects of quarantine and isolation on the number of hospitalized cases.  
\section{Introduction}
Many existing studies focus on the impact of quarantine or isolation of exposed and infectious individuals on COVID-19 dynamics \cite{gu2022mathematical,huang2021analysis,memon2021assessing}. 
Isolation is the separation of ill people  with contagious diseases from non-infected individuals  to protect non-infected individuals. Quarantine is one of the oldest and most effective tools for controlling communicable disease outbreaks. It involves restriction, to the home or a designated facility, of individuals presumed to have been exposed to a contagious disease but are not ill, either because they did not become infected or because they are still in the incubation period. It may be applied at the individual or group level and may be voluntary or mandatory \cite{wilder2020isolation}.\\
Calistus N. Ngonghala et al. \cite{ngonghala2020mathematical} developed a  mathematical model for studying the transmission dynamics and control of the COVID-19 pandemic in the state of New
York, US. The model takes the form of a
Kermack McKendrick, compartmental, deterministic system of non-linear differential equations \cite{kermack1927contribution}. It incorporates features pertinent to
COVID-19 transmission dynamics and control, such as the quarantine of
suspected cases and the isolation/hospitalization of confirmed COVID-19 cases. The model is parametrized using available COVID-19 mortality data
to assess the population-level impact of the main intervention strategies being implemented in the state (in particular, quarantine, isolation, contact tracing, social distancing and the use of face masks in public). Their study shows that wide-scale implementation of quarantine
intervention may not be very effective (in minimizing the burden of
COVID-19) if the strategy of isolating confirmed cases is effective. In
other words, their study suggests that if isolation can be implemented
effectively (high efficacy and coverage), then quarantine of people
suspected of contracting COVID-19 may not be necessary. The study emphasizes the vital role social distancing plays in curtailing the burden of COVID-19. Using face masks in public is very useful in minimizing community transmission, and the burden of COVID-19 provided their coverage level is high.\\
Renquan Zhang, Yu Wang, Zheng Lv and Sen Pei \cite{zhang2022evaluating} developed a mathematical model to estimate the effect of quarantine
on suppressing COVID-19 spread in four cities: Wuhan in China, New York City in the US, Milan in Italy and London in the UK. They incorporated a component of quarantine into a classical susceptible-exposed-infected-removed (SEIR) model, and calibrated the model to confirmed cases in each city during the early phase of
the pandemic using a data assimilation method. They estimated key epidemiological parameters before lockdown in each city, and evaluated the impact of the quarantine rates of susceptible, exposed and undetected infected populations on disease transmission. Particularly,
they estimated the required minimal quarantine rates of those populations to reduce the effective reproductive number below one at the beginning of lockdown.\\
\\
There are few studies involving mathematical models to study the COVID-19 situation in Oman. Abraham Varghese et al.\cite{varghese2021seamhcrd} developed a mathematical model to analyze the pandemic's nature using Oman's data. The model is an extension of SEIR where they expanded the infected compartments into mild, moderate, severe and critical, based on the clinical stages of infection. The parameters were estimated by fitting the data of Oman to the model differential equations. They justified the estimated parameter 
values  with the effective actions taken by the government to lessen the spread of the disease. More studies about COVID-19 in Oman are found in \cite{ aloptimal, ebraheem2021delayed,mahmoud2020forecasting,zia2020covid}.\\
The increasing number of infections, especially with severe symptoms that require medical care and hospitalization, undoubtedly whelms the health systems. It may result in increased death cases not only directly because of COVID-19, but also because persons with other medical conditions may not get the needed care of hospitalization or follow-up when the health systems become over pressured. Moreover, when many cases need hospitalization, it becomes difficult to find designated places to isolate them without affecting other workers. Therefore, the risk of transmission from this group and infecting the front-liners become higher. Hence, when many of health workers get infected and, accordingly, isolated, this situation will impact the quality of the medical care with the reduced number of caregivers. \\ 
This current study suggests an epidemic mathematical model to investigate and explore the relationship between the quarantine rate of healthy people and the  isolation rate of infectious population from one side and the registered hospitalized cases in Oman from the other side. It is important to explore such relations to limit and reduce COVID-19 cases that require medical aid so that other patients suffering from other illnesses still get the same level and chance of the needed medical care  and will not become affected by COVID-19's high demands.
This paper is organized as follows. The model formulation is given in the next section. Section 3 is devoted for model analysis including non-negativity of solution, stability analysis and reproduction number.  Fitting of the model and numerical simulation are presented in Section 4 to illustrate the effects of  parameters related to quarantine and isolation to health care system. Finally, the conclusion is given in Section 5.
\section{Model Formulation}
The proposed model is formulated based on sub-dividing the human population, $N(t)$, into ten compartments of susceptible $S(t)$,  exposed $E(t)$,  quarantined  $Q(t)$, asymptomatically-infectious $A(t)$, pre-symptomatically-infectious $P(t)$, symptomatically-infectious $I(t)$, isolated $J(t)$, hospitalized $H(t)$, hospitalized in the intensive care unit (ICU) $C(t)$ and recovered $R(t)$, so that
$$
N(t) = S(t) +  E(t) + Q(t) + A(t) + P(t) + I(t) + J(t) + H(t) + C(t) + R(t)
$$
\subsection{Model Assumptions and Description}
Here, we give the assumptions and description of the model. We assume that $ S(t)$ are the susceptible individuals who are at risk of getting infected by the COVID-19 virus. They become infected with COVID-19 by effective contact with asymptomatic, presymptomatic, symptomatic, isolated or hospitalized individuals at rates of $\beta_{{A}}$, $\beta_{{P}}$, $\beta_{{I}}$, $\beta_{{J}}$ and $\beta_{{H}}$, respectively. It is assumed that transmission from individuals in the ICU is negligible due to the complete isolation of this group. Moreover, it is assumed that the commitment level of people in the isolation class varies. Therefore, a portion $\varepsilon_J$ of non-adherent isolated individuals accounts for transmitting the disease.\\
\\
Some susceptible individuals are put under quarantine, $Q(t)$,  at a rate of $\rho_{{S}}$ and leave the class by returning to the susceptible class at a rate of $\rho_{{Q}}$. People usually practice quarantine due to either personal behaviours of awareness/fear or  compliance to governmental intervention measures. These measures include lockdown strategy, closures of schools and non-essential businesses, and cancelling large public and social gathering events. We assume total protection of individuals in this class, and no infection can occur. \\ 
\\
All newly infected individuals enter the exposed class $E(t)$. People in this class are considered infected but not yet infectious during their latent period. After this period, they become infectious at different rates of $\lambda_A, \lambda_P$ and $ \lambda_I$ as asymptomatic, presymptomatic and symptomatic, respectively.\\
\\
Asymptomatic individuals, $A(t)$, are infectious and contagious but do not show symptoms until they recover at a rate of $\gamma_A$. People show no clinical symptoms because the viral load in their bodies is very small. Consequently, the transmission rate from this class is very low compared to others with apparent symptoms. \\ 
\\
Presymptomatic infectees, $P(t)$, currently have no symptoms  but will eventually show symptoms after some time. They stay in their class until they develop symptoms at a rate of $\sigma$ and move to the symptomatic class. This means that, on average, an infected individual takes $\D\frac{1}{\sigma}$ days from being infectious to developing symptoms and moving to the next stage of the disease.\\
\\
Individuals in the symptomatic class, $I(t)$, are those with mild COVID-19 symptoms. Depending on their identification and symptoms severity, some become isolated at  rate of $\varepsilon$ and some move to hospitals for medical care at a rate of $\psi_I$. The remaining stay in the symptomatic class until they recover at a rate of $\gamma_{I}$. No disease-related deaths occur in this class. \\ 
\\
The infected isolated group, $J(t)$, contains confirmed cases of COVID-19 who are ordered to isolate themselves at home or institutionally. Some leave this class at a rate of $\psi_J$ for medical care when their health status becomes more severe, while others stay at their class until they recover at a rate of $\gamma_J$. We assume that a portion of $\varepsilon_J$ of isolated people, especially at home isolation, is not fully adherent to isolation measures and spreads the disease.\\
\\
Patients in the hospitalized class, $H(t)$, are those with severe COVID-19 symptoms. If their symptoms become even more severe, they are transferred to the ICU at a rate of $\psi_H$. Otherwise, they stay until they recover at a rate of $\gamma_{H}$.\\
\\
The ICU patients, $C(t)$, have the worst symptoms. When the medical intensive care is successful, some will recover and are discharged at a rate of $\gamma_C$. However, some patients do not respond even with intensive care and die due to the disease at a rate of $\delta_C$.\\
\\
We assume that there are no human demographic processes of births, immigration or deaths caused by other means than COVID-19. Such disease transmission models are called \textit{epidemic models}\cite{kermack1927contribution}.
\subsection{Model Equations}
\begin{figure}[H]
	 \includegraphics[scale=0.55,center]{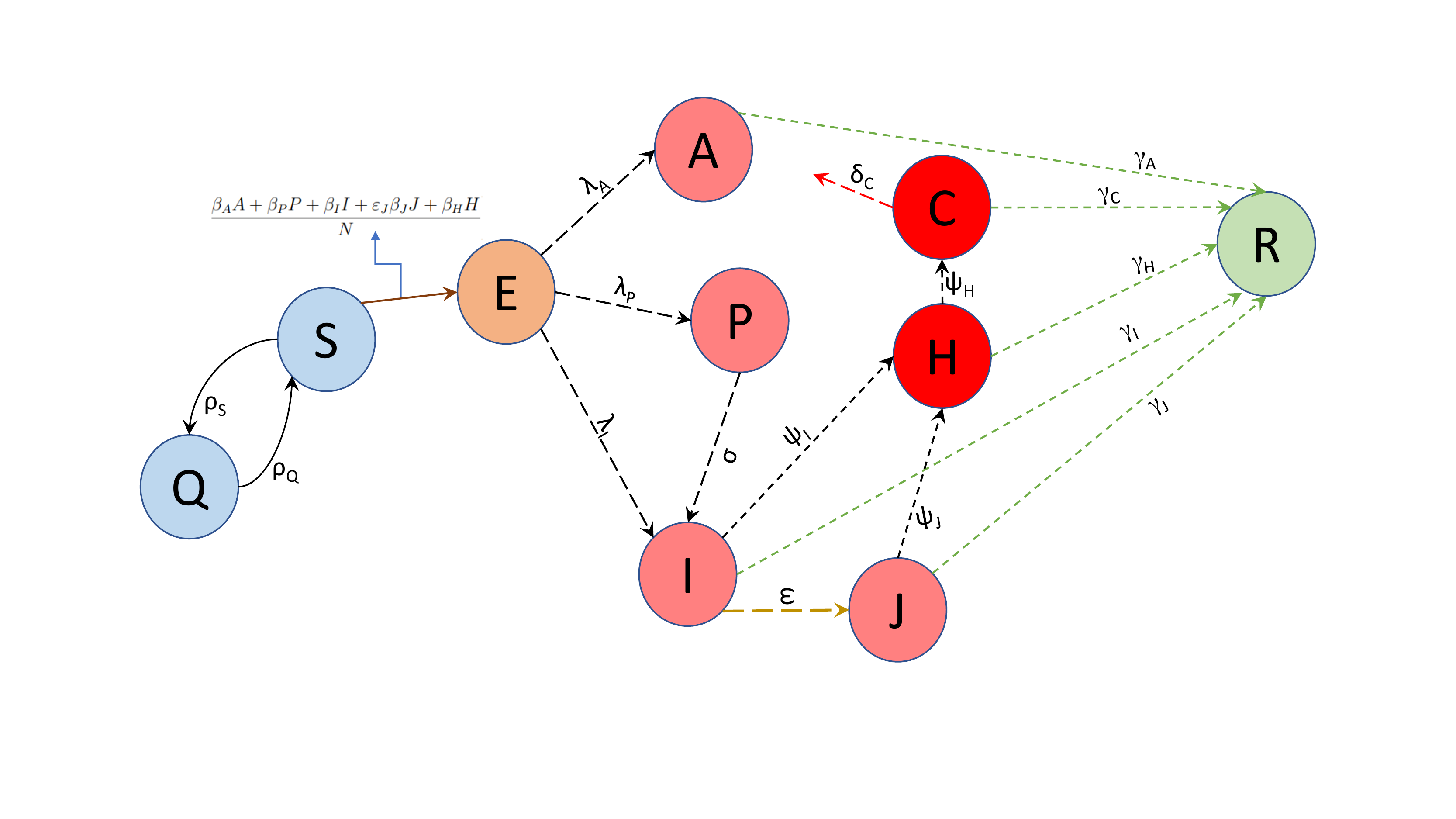}
	\caption{Model (\ref{model:1}) compartments flow chart.}
	\label{Fig: flowchart}
\end{figure}

Based on the above description, the proposed mathematical model is given by the following system of non-linear differential equations:
\begin{eqnarray} 
\dfrac{dS}{dt} &=&   \rho_Q Q  -  \left[\dfrac{\beta_A A + \beta_P P + \beta_I I +  \varepsilon_J \beta_J J  + \beta_H H }{N  } \right]S  -   \rho_S   S,  \nonumber \vspace{2mm}  \nonumber \\
\nonumber\\[8pt] %
\dfrac{dE}{dt} &=&     \left[\dfrac{\beta_A A + \beta_P P + \beta_I I +  \varepsilon_J \beta_J J  + \beta_H H }{N  } \right]S  - ( \lambda_A + \lambda_P + \lambda_I ) E, \nonumber \\
\dfrac{dQ}{dt} &=& \rho_S S -  \rho_Q  Q, \nonumber
\nonumber \vspace{2mm}\\[8pt] 
\dfrac{dA}{dt} &=& \lambda_A E - \gamma_A  A, \nonumber  \vspace{2mm}\\[8pt] 
\dfrac{dP}{dt} &=& \lambda_P E -   \sigma P,  \label{model:1} \vspace{2mm}\\[8pt] \label{model} 
\dfrac{dI}{dt} &=& \lambda_I E +  \sigma P  - ( \varepsilon+ \psi_I+ \gamma_{I} ) I,   \nonumber \vspace{2mm}  \\[8pt]  
\dfrac{dJ}{dt} &=& \varepsilon I   - ( \psi_J+ \gamma_{J} ) J,   \nonumber \vspace{2mm} \\[8pt]
\dfrac{d H}{dt} &=& \psi_I I + \psi_J J   -(\psi_H +\gamma_H ) H, \nonumber \vspace{2mm}\\[8pt] 
\dfrac{d C}{dt} &=& \psi_H H  -(\gamma_C + \delta_C ) C, \nonumber \vspace{2mm}\\[8pt]
\dfrac{dR}{dt} &=& \gamma_A A + \gamma_{I} I + \gamma_{J} J + \gamma_{H} H + \gamma_C C,  \nonumber  
\end{eqnarray}  
\\ 
where, $N'(t)=  - \delta_C C(t)$.\\
\\
A flow diagram of the model is shown in Fig. \ref{Fig: flowchart}. Parameters of the model are described in Table \ref{table1}.
\noindent For analysis purposes, we consider the equation for the rate of change of the deceased individuals given by
\begin{align*}
\dfrac{d D}{dt} &=  \delta_C  C(t).
\end{align*}
Specifically, this equation will be used in fitting the model equations to the reported death data. The population size, $N(t)$, is not constant, and we assume it is decreasing due to the COVID-19-related deaths.
\begin{table}[ht]
	\caption{Parameters used in the model (\ref{model:1}).}
	\label{table1}
	\centering
	\begin{tabular}{|c| p{10cm}| }
		\hline \hline
		Parameter  & Description  \\ \hline
	    $\beta_A$ & Transmission rate  from asymptomatic individuals.   \\ 
		$\beta_P$ & Transmission rate from  presymptomatic individuals.  \\ 
		$\beta_I$ & Transmission rate  from  symptomatic individuals.   \\
		$\beta_J$ & Transmission rate from  isolated individuals. \\ 
		$\beta_H$ & Transmission rate  from  hospitalized individuals. \\ 
		$\varepsilon_J$ & Portion of isolated who transmit the disease. \\
		$\rho_S$ &  Quarantine rate of susceptible.  \\
		$\rho_Q$ & Rate of leaving  quarantine class.  \\
		$\lambda_A$ & Rate at which exposed becomes asymptomatic.  \\
		$\lambda_P$ & Rate at which exposed becomes presymptomatic. \\
		$\lambda_I$ & Rate at which exposed becomes infected with  symptoms. \\
		$\sigma$ & Rate at which presymptomatic develops symptoms. \\
		$\varepsilon$ & Rate of isolation of symptomatic individuals.   \\
		$\psi_I$ & Hospitalization rate of symptomatic patients. \\
		$\psi_J$ & Hospitalization rate of isolated individuals.  \\ 
		$\psi_H$ & Rate at which hospitalized individual is transferred to the ICU.  \\
		$\gamma_A$ & Recovery rate of asymptomatic.  \\
		$\gamma_I$ & Recovery rate of symptomatic. \\
		$\gamma_J$ & Recovery rate of isolated.  \\
		$\gamma_H$ & Discharge rate of hospitalized patients.   \\
		$\gamma_C$ & Discharge rate of ICU patients.  \\
		$\delta_C$ & COVID-19 related death rate of the ICU class.   \\
		\hline \hline
	\end{tabular}
\end{table}
\newpage
\section{Model Analysis}
\subsection{Positivity of Solutions}
The analysis of the model is carried out in the feasible region:
$$\varOmega = \lbrace \left( S, E, Q, A, P, I, J, H, C, R\right) \in \mathbb{R}_+^{10}: S+E+Q+A+P+I+J+H+C+R \le N(0)  \rbrace,$$
where $N(0)$ is the initial total population size.
\begin{Thm}
The region $\varOmega$ is positively-invariant and attracts all solutions of the  model (\ref{model:1}) for all $t>0.$ 
\end{Thm} 
\noindent \textit{Proof} Let $\textbf{U}(0)=\left(S(0),E(0),Q(0), A(0), P(0), I(0), J(0), H(0), C(0), R(0)\right)$ be the initial condition of the system (\ref{model:1}). We need to prove that solutions starting in $\varOmega$, with initial conditions $\textbf{U}(0)\ge 0$ remain in $\varOmega$ for all $t>0$.\\ Let $t_1 = \text{sup}\left( t > 0 \mid \textbf{U}(t)>0  \right)$. \\
Consider the first equation of model (\ref{model:1}):
\begin{align*}
\dfrac{dS}{dt} &=&   \rho_Q Q  -  \left[\dfrac{\beta_A A + \beta_P P + \beta_I I +  \varepsilon_J \beta_J J  + \beta_H H }{N  } \right]S  -   \rho_S   S, 
\end{align*}
and rewrite the force of infection term in the form
\begin{align*}
\left[\dfrac{\beta_A A + \beta_P P + \beta_I I +  \varepsilon_J \beta_J J  + \beta_H H }{N  } \right]S = \beta(t) S.
\end{align*} 
Then, we have\\ 
\\
$\dfrac{dS}{dt} +  \left(\beta(t) + \rho_S \right) S =  \rho_Q Q.$\\   Multiplying by the integrating factor  $ e^{\D\int_{0}^{t} \beta(v)dv + \rho_S t}$,  and integrating both sides over $(0,  t_1)$  gives
\begin{align*}
&S(t_1) e^{\D\int_{0}^{t_1} \beta(v) dv + \rho_S t_1} - S(0)  = \rho_Q \D\int_{0}^{t_1}Q(t) \,e^{\D\int_{0}^{t} \beta(v)dv + \rho_S t} dt  \\
\\
&S(t_1) = \left[  S(0) + \rho_Q \D\int_{0}^{t_1} Q(t) e^{\D\int_{0}^{t} \beta(v)dv + \rho_S t} dt \right] \,e^{-\D\int_{0}^{t_1} \beta(v)dv + \rho_S t_1} > 0.
\end{align*}
Similarly, one can establish that other components are positive at $t=t_1$. Hence, using continuity of solutions and $\textbf{U}(0)\ge 0$, $t_1$ cannot be a supremum. Therefore, all the solutions remain positive for all time $t> 0$.
\subsection{Stability Analysis of Continuum of Disease Free Equilibria and Reproduction Number}
The Model has a continuum of disease-free equilibria (DFE), given by
$$\mathcal{E}_0= \left( S^*, E^*, Q^*, A^*, P^*, I^*, J^*, H^*, C^*, R^*\right)=\left( S^*, 0, Q^*, 0, 0, 0, 0, 0, 0, R^*\right), $$
where $S^*+Q^*+R^*=N^*$ is the total population size at the disease free state.\\
Moreover, from the first and third equations of the model (\ref{model:1}), at the DFE, $Q^*=\D\frac{\rho_S}{\rho_Q}S^*$, and $S^*+Q^*=\left(\D\frac{\rho_Q + \rho_S}{\rho_Q} \right)S^*$.\\
Therefore, the continuum of the DFE can be written in the form
$$\mathcal{E}_0=\left( \D\frac{\rho_Q }{\rho_Q+ \rho_S}\left(N^*-R^*\right), 0, \D\frac{\rho_S }{\rho_Q+ \rho_S}\left(N^*-R^*\right), 0, 0, 0, 0, 0, 0, R^*\right).$$
Next generation method, described in \cite{diekmann1990definition}, is used to derive the reproduction number.\\
The matrix of new infections is
\begin{align}
\mathcal{F}= \left[ \begin {array}{c} {\D\frac { \left( \beta_{{A}}A+\beta_{{P}}P+
		\beta_{{I}}I+\varepsilon_{{J}}\beta_{{J}}J+\beta_{{H}}H \right) S}{N}}
\\ \noalign{\medskip}0\\ \noalign{\medskip}0\\ \noalign{\medskip}0
\\ \noalign{\medskip}0\\ \noalign{\medskip}0\\ \noalign{\medskip}0
\end {array} \right],
\end{align}
and the transition matrix is
\begin{align}
\mathcal{V}=\left[ \begin {array}{c}  \left( \lambda_{{A}}+\lambda_{{P}}+\lambda_
{{I}} \right) E\\ \noalign{\medskip}-\lambda_{{A}}E+\gamma_{{A}}A
\\ \noalign{\medskip}-\lambda_{{P}}E+\sigma\,P\\ \noalign{\medskip}-
\lambda_{{I}}E-\sigma\,P+ \left( \varepsilon+\psi_{{I}}+\gamma_{{I}}
\right) I\\ \noalign{\medskip}-\varepsilon\,I+ \left( \psi_{{J}}+\gamma_
{{J}} \right) J\\ \noalign{\medskip}-\psi_{{I}}I-\psi_{{J}}J+ \left( 
\psi_{{H}}+\gamma_{{H}} \right) H\\ \noalign{\medskip}-\psi_{{H}}H+
\left( \gamma_{{C}}+\delta_{{C}} \right) C\end {array} \right].
\end{align}
The Jacobian of the matrix of infection evaluated at the DFE is
\begin{align}
F&=\left[ \begin {array}{ccccccc} 0&{\D\frac {\beta_{{A}}S^*}{N^*}}&{\D\frac {\beta_{{P}}S^*}{N^*}}&{\D\frac {\beta_{{I}}S^*}{N^*}}&{\D\frac {
		\varepsilon_{{J}}\beta_{{J}}S^*}{N^*}}&{\D\frac {
		\beta_{{H}}S^*}{N^*}}&0\\ \noalign{\medskip}0
&0&0&0&0&0&0\\ \noalign{\medskip}0&0&0&0&0&0&0\\ \noalign{\medskip}0&0
&0&0&0&0&0\\ \noalign{\medskip}0&0&0&0&0&0&0\\ \noalign{\medskip}0&0&0
&0&0&0&0\\ \noalign{\medskip}0&0&0&0&0&0&0\end {array} \right], 
\end{align} 
and the Jacobian of the transition matrix is
	\begin{align}
	V=\left[ \scalemath{0.91}{\begin {array}{ccccccc} \lambda_{{A}}+\lambda_{{P}}+\lambda_{{
		I}}&0&0&0&0&0&0\\ \noalign{\medskip}-\lambda_{{A}}&\gamma_{{A}}&0&0&0&0
&0\\ \noalign{\medskip}-\lambda_{{P}}&0&\sigma&0&0&0&0
\\ \noalign{\medskip}-\lambda_{{I}}&0&-\sigma&\varepsilon+\psi_{{I}}+
\gamma_{{I}}&0&0&0\\ \noalign{\medskip}0&0&0&-\varepsilon&\psi_{{J}}+
\gamma_{{S}}&0&0\\ \noalign{\medskip}0&0&0&-\psi_{{I}}&-\psi_{{J}}&
\psi_{{H}}+\gamma_{{H}}&0\\ \noalign{\medskip}0&0&0&0&0&-\psi_{{H}}&
\gamma_{{C}}+\delta_{{C}}\end {array}} \right] .
\end{align}
The next generation matrix (NGM) is\\
\\
$FV^{-1}=\left[ \begin {array}{ccccccc} R_C&{\D\frac {\beta_{{A}}S^*}{  \gamma_{{A}}N^*}}&a_{13}&a_{14}&a_{15}&{
	\D\frac {\beta_{{H}}S^*}{ \left( \psi_{{H}}+\gamma_{{H}} \right) N^* }}&0\\ \noalign{\medskip}0&0&0
&0&0&0&0\\ \noalign{\medskip}0&0&0&0&0&0&0\\ \noalign{\medskip}0&0&0&0
&0&0&0\\ \noalign{\medskip}0&0&0&0&0&0&0\\ \noalign{\medskip}0&0&0&0&0
&0&0\\ \noalign{\medskip}0&0&0&0&0&0&0\end {array} \right],
$
\newpage
where \\
\begin{dmath*}
a_{13}=\left[{\D\frac {\beta_{{P}}}{
		\sigma }}+{\D\frac {\beta_{{I}}}{   \left( \varepsilon+\psi_{{
				I}}+\gamma_{{I}} \right) }}+{\D\frac {\varepsilon_{{J}}\beta_{{J}}\varepsilon }{   \left( \psi_{{J}}+
		\gamma_{{J}} \right)  \left( \varepsilon+\psi_{{I}}+\gamma_{{I}} \right)  
}}+ {\D\frac {\beta_{{H}} \left( \psi_{{I}} \left( \psi_{{J}}+
		\gamma_{{J}} \right) +\psi_{{J}}\varepsilon \right) }{\left( \psi_{{J}}+\gamma_{{J}} \right)  \left( 
		\varepsilon+\psi_{{I}}+\gamma_{{I}} \right)  \left( \psi_{{H}}+\gamma_{{H
		}} \right)}}\right]\D\frac{S^*}{N^*},\\ \\
 a_{14}=\left[{\D\frac {\beta_{{I}}}{ \left( \varepsilon+\psi_{{I}}+\gamma_{{I}} \right) }}+{
		\D\frac {\varepsilon_{{J}}\beta_{{J}}\varepsilon }{ \left( \psi_{{J}}+\gamma_{{J}} \right)  \left( 
			\varepsilon+\psi_{{I}}+\gamma_{{I}} \right) }}+
		{\D\frac {\beta_{{H}} \left( \psi_{{I}} \left( \psi_{{J}}+\gamma_{{J}} \right) +\psi_{{J
			}}\varepsilon \right) }{  \left( 
			\psi_{{J}}+\gamma_{{J}} \right)  \left( \varepsilon+\psi_{{I}}+\gamma_{{I
			}} \right)  \left( \psi_{{H}}+\gamma_{{H}} \right) }}\right]\D\frac{S^*}{N^*},\\
		\\ \\
a_{15}=\left[{\D\frac {\varepsilon
		_{{J}}\beta_{{J}}}{ 
		\left( \psi_{{J}}+\gamma_{{J}} \right) }}+{\D\frac {\beta_{{H}}\psi_{{J}}}{ \left( \psi_{{J}}
		+\gamma_{{J}} \right)  \left( \psi_{{H}}+\gamma_{{H}} \right)  }}\right]\D\frac{S^*}{N^*} \quad \text{ and}\\
\end{dmath*}
$R_C= \rho\left(FV^{-1}\right)=R_A + R_P + R_I + R_J + R_H,\\
\\
\text{where}\\
\\
R_A={\D\frac {\beta_{{A}}\lambda_{{A}}S^*}{\left( \lambda_{{A}}+\lambda_{{P}}+\lambda_{{I}} \right) 
		\gamma_{{A}}N^*}}, \: \: \qquad \qquad
R_P={\D\frac {\beta_{{P}}\lambda_{{P}}S^*}{\left( \lambda_{{A}}+\lambda_{{P}}+
		\lambda_{{I}} \right) \sigma N^*}},\\
	\\ \\
R_I={\D\frac {\beta_{{I}} \left( 
		\lambda_{{I}}+\lambda_{{P}} \right) S^* }{   \left( \lambda_{{A}}+\lambda_{{P}}+\lambda_{{I}} \right) 
		\left( \varepsilon+\psi_{{I}}+\gamma_{{I}} \right) N^* }}, \\
	\\ \\
R_J={\D\frac {\varepsilon_{
			{J}}\beta_{{J}}\varepsilon\, \left( \lambda_{{I}}+\lambda_{{P}}
		\right) S^* }{ \left( \lambda_{{A}}
		+\lambda_{{P}}+\lambda_{{I}} \right)  \left( \varepsilon+\psi_{{I}}+
		\gamma_{{I}} \right)  \left( \psi_{{J}}+\gamma_{{J}} \right) N^* }},\\
	\\ \\
R_H={
	\D\frac {\beta_{{H}} \left( \lambda_{{I}}+\lambda_{{P}}
		\right)  \left( \psi_{{I}} \left( \psi_{{J}}+\gamma_{{J}} \right) +
		\psi_{{J}}\varepsilon \right)S^* }{ 
		\left( \lambda_{{A}}+\lambda_{{P}}+\lambda_{{I}} \right)  \left( 
		\varepsilon+\psi_{{I}}+\gamma_{{I}} \right)  \left( \psi_{{J}}+\gamma_{{J
		}} \right)  \left( \psi_{{H}}+\gamma_{{H}} \right) N^* }}.	
	$\\
\\	
The quantity $R_C$ is called the control reproduction number of the model (\ref{model:1}). It measures the average number of new infected cases when an infectious individual is introduced to a population where a control strategy is adopted. In our current proposed model, quarantine and isolation are the adopted control strategies. $R_C$ is the sum of the five constituent reproduction numbers associated with the number of new cases produced by asymptomatic $(R_A)$, presymptomatic $(R_P)$, symptomatic $(R_I)$, isolated $(R_J)$ and hospitalized $(R_H)$ individuals.\\	
The upper bound of the reproductive number occurs when $R^*=0$. In this case, the total population at the equilibrium state, $N^*$, becomes: $N^*=S^*+Q^*= \left(\D\frac{\rho_Q + \rho_S}{\rho_Q}\right)S^*$, and $R_C$ becomes:\\
\begin{dmath}
	$$ R_C=\D\frac{\rho_Q}{\left(\rho_{{Q}}+\rho_{{S}} \right)\left( \lambda_{{A}}+\lambda_{{P}}+\lambda_{{I}} \right)}\left[ {\D\frac {\beta_{{A}}\lambda_{{A}}}{   
		\gamma_{{A}}}} + \D\frac{\beta_P \lambda_P}{\sigma} + \D\frac{\beta_I \left( \lambda_I + \lambda_P\right)}{\left( \varepsilon+\psi_{{I}}+\gamma_{{I}} \right)} + \D\frac{\varepsilon\,\varepsilon_{
		{J}}\beta_{{J}} \left( \lambda_{{I}}+\lambda_{{P}}
	\right) }{\left( \varepsilon+\psi_{{I}}+ 
	\gamma_{{I}} \right)  \left( \psi_{{J}}+\gamma_{{J}} \right)} +
 \D\frac{\beta_{{H}} \left( \lambda_{{I}}+\lambda_{{P}}
	\right)  \left( \psi_{{I}} \left( \psi_{{J}}+\gamma_{{J}} \right) +
	\psi_{{J}}\varepsilon \right)}{\left( 
	\varepsilon+\psi_{{I}}+\gamma_{{I}} \right)  \left( \psi_{{J}}+\gamma_{{J
	}} \right)  \left( \psi_{{H}}+\gamma_{{H}} \right)} \right]. $$ \label{R_C formula}
\end{dmath}
The result below follows from theorem 2 of \cite{van2002reproduction}.
\begin{Thm} \label{thm: local stb}
	The continuum of the disease free equilibria ($\mathcal{E}_0$) is locally asymptotically stable if $R_C < 1$. If $R_C>1$, the epidemic rises to a peak and then eventually declines to zero.
\end{Thm}
\noindent From an epidemiological sense, Theorem \ref{thm: local stb} implies that a few cases of COVID-19 will not result in an outbreak in the community if $R_C<1$. This means that the disease will die out rapidly when $R_C<1$ if the initial number of infected individuals is in the basin of attraction of the continuum of the DFE. In fact, for epidemic models with no demographic dynamics, the condition $R_C<1$ is sufficient but not necessary for eliminating the epidemic. Even when $R_C>1$, the epidemic eventually dies out with time due to many factors such as control measures \cite{gumel2021primer,iboi2020mathematical}.\\
The above theorem requires that the initial size of the sub-populations have to be within the basin of attraction of the DFE to eliminate the disease when $R_C<1$. Therefore, for the disease elimination to be independent of the initial size, we must show that the DFE is globally-asymptotically stable.
\begin{Thm} \label{thm: global stb}
     The continuum of the disease free equilibria ($\mathcal{E}_0$) of the model (\ref{model:1}) is globally-asymptotically stable in $\varOmega$ if $R_C \le 1$.
\end{Thm}
\textit{Proof.} The proof is based on using Lyapunov function theory. Consider the model (\ref{model:1}) with the following Lyapunov function:
\begin{align*} 
\mathcal{L} &= E + g_1 A + g_2P + g_3 I + g_4 J + g_5H,\\
\text{where, } g_1&=\D\frac{\beta_A }{\gamma_A }, \qquad g_2= g_3 + \D\frac{\beta_P}{\sigma },\\
g_3&=\D\frac{1}{\left(\varepsilon + \psi_I + \gamma_I\right)} \left[\beta_I + g_4 \varepsilon  + g_5 \psi_I\right],\\
g_4&= \D\frac{ 1}{\left(\psi_J + \gamma_J\right)} \left(g_5 \psi_J  + \varepsilon_J \beta_J\right) \text{ and } g_5= \D\frac{\beta_H}{\left(\psi_H + \gamma_H\right)}. \intertext{ It follows that the Lyapunov derivative is given by}
\mathcal{\dot{L}} &= \dot{E} + g_1 \dot{A} + g_2\dot{P} + g_3 \dot{I} + g_4 \dot{J} + g_5\dot{H},\\
&= \left[\left(\beta_A A + \beta_P P + \beta_I I +  \varepsilon_J \beta_J J  + \beta_H H \right) \D\frac{S}{N} - ( \lambda_A + \lambda_P + \lambda_I ) E\right] \\
&+ g_1 \left[ \lambda_A E - \gamma_A  A \right] +g_2 \left[ \lambda_P E -   \sigma P\right]\\
&+ g_3 \left[\lambda_I E +  \sigma P  - ( \varepsilon+ \psi_I+ \gamma_{I} ) I\right]\\
& + g_4\left[  \varepsilon I    -(\psi_J +\gamma_J ) J \right] + g_5 \left[ \psi_I I + \psi_J J   -(\psi_H +\gamma_H ) H \right],
\intertext{which can be simplified to,}
\mathcal{\dot{L}} &=\left( \beta_A \D\frac{S}{N} -g_1\gamma_A\right)A + \left( \beta_P \D\frac{S}{N} + g_3 \sigma - g_2 \sigma \right)P \\
&+ \left( \beta_I \D\frac{S}{N} + g_4 \varepsilon + g_5 \psi_I - g_3 (\varepsilon + \psi_I + \gamma_I) \right)I + \left( \varepsilon_J \beta_J \D\frac{S}{N} +g_5 \psi_J -g_4\left( \psi_J + \gamma_J\right)\right)J\\
&+ \left(\beta_H \D\frac{S}{N} - g_5\left( \psi_H + \gamma_H\right)\right)H 
+\left( g_1\lambda_A + g_2 \lambda_P + g_3\lambda_I\right)E - \left(\lambda_A+\lambda_P+\lambda_I\right)E,\\
\\
&= \beta_A\left(\D\frac{S}{N} -1\right)A + \beta_P\left(\D\frac{S}{N} -1\right)P + \beta_I\left(\D\frac{S}{N} -1\right)I + \varepsilon_J\beta_J\left(\D\frac{S}{N} -1\right)J \\
&+ \beta_H\left(\D\frac{S}{N} -1\right)H + \left( g_1\lambda_A + g_2 \lambda_P + g_3\lambda_I\right)E - \left(\lambda_A+\lambda_P+\lambda_I\right)E.
\end{align*}
Noting that $\D\frac{S}{N} \le 1$, 
\begin{dmath*} 
\mathcal{\dot{L}} \le \left( g_1\lambda_A + g_2 \lambda_P + g_3\lambda_I\right)E - \left(\lambda_A+\lambda_P+\lambda_I\right)E,\\
= \left(\lambda_A+\lambda_P+\lambda_I\right)\left[\D\frac{1}{\left(\lambda_A+\lambda_P+\lambda_I \right)}\left(g_1 \lambda_A + g_2 \lambda_P +g_3 \lambda_I\right) -1 \right] E,
= \left(\lambda_A+\lambda_P+\lambda_I\right)\left[\D\frac{1}{\left(\lambda_A+\lambda_P+\lambda_I \right)}\left(\D\frac{\beta_A \lambda_A}{\gamma_A} + \D\frac{\beta_P \lambda_P}{\sigma} +g_3 \left(\lambda_I +\lambda_P\right) \right) -1 \right] E,
= \D\frac{\left(\rho_Q + \rho_S\right)\left(\lambda_A+\lambda_P+\lambda_I\right)}{\rho_Q} \left[\D\frac{\rho_Q}{\left( \rho_Q + \rho_S\right)\left(\lambda_A+\lambda_P+\lambda_I \right)} \left( \D\frac{\beta_A \lambda_A}{\gamma_A} + \D\frac{\beta_P \lambda_P}{\sigma} + \D\frac{\beta_I \left( \lambda_I + \lambda_P\right)}{\left( \varepsilon+\psi_{{I}}+\gamma_{{I}} \right)} + \D\frac{\varepsilon\,\varepsilon_{
		{J}}\beta_{{J}} \left( \lambda_{{I}}+\lambda_{{P}}
	\right) }{\left( \varepsilon+\psi_{{I}}+ 
	\gamma_{{I}} \right)  \left( \psi_{{J}}+\gamma_{{J}} \right)} +
\D\frac{\beta_{{H}} \left( \lambda_{{I}}+\lambda_{{P}}
	\right)  \left( \psi_{{I}} \left( \psi_{{J}}+\gamma_{{J}} \right) +
	\psi_{{J}}\varepsilon \right)}{\left( 
	\varepsilon+\psi_{{I}}+\gamma_{{I}} \right)  \left( \psi_{{J}}+\gamma_{{J
	}} \right)  \left( \psi_{{H}}+\gamma_{{H}} \right)}  \right)-1\right]E
=\D\frac{\left(\rho_Q + \rho_S\right)\left(\lambda_A+\lambda_P+\lambda_I\right)}{\rho_Q}\left[R_C - 1\right]E.
	\end{dmath*}
Therefore, $\mathcal{\dot{L}} \le 0$ if $R_C\le 1$, and $\mathcal{\dot{L}} =0$ if and only if $E(t) =0$. Substituting $E(t)=0$ in the model (\ref{model:1}) shows that the solution\\ $ \left( S(t), E(t), Q(t), A(t), P(t), I(t), J(t), H(t), C(t), R(t)\right) \rightarrow \left( S^*, 0, Q^*, 0, 0, 0, 0, 0, 0, R^*\right),$ as $t \rightarrow \infty$. Moreover, it can be shown that the largest compact invariant set in $\left\lbrace \left( S(t), E(t), Q(t), A(t), P(t), I(t), J(t), H(t), C(t), R(t)\right) \in \varOmega : \mathcal{\dot{L}} =0 \right\rbrace$ is the continuum of disease free equilibria $(\mathcal{E}_0)$. Thus, by LaSalle's Invariance Principle \cite{lasalle}, the disease free equilibria of the model (\ref{model:1}) is globally-asymptotically stable in $\varOmega$ whenever $R_C\le 1$.\\
\\
\textbf{The Basic reproduction number:} When no control strategy is adopted, especially at the beginning of the pandemic, the quantity used to measure the new cases generated by a single infected case in a completely susceptible population is called the basic reproduction number\cite{diekmann1990definition} $R_0$. In this sense, $R_0$ is the measure of the worst-case scenario of the disease when no health control or intervention measures are implemented in the community. In our model, quarantine and isolation are the proposed control methods. Therefore,  model (\ref{model:1}) reduces to the below form without control.
\begin{eqnarray} 
\dfrac{dS}{dt} &=&   -  \left[\dfrac{\beta_A A + \beta_P P + \beta_I I +  \beta_H H }{N  } \right]S ,  \nonumber \vspace{2mm}  \nonumber \\
\nonumber\\[8pt] %
\dfrac{dE}{dt} &=&     \left[\dfrac{\beta_A A + \beta_P P + \beta_I I +   \beta_H H }{N  } \right]S  - ( \lambda_A + \lambda_P + \lambda_I ) E, \nonumber \\
\nonumber \vspace{2mm}\\[8pt]  
\dfrac{dA}{dt} &=& \lambda_A E - \gamma_A  A, \nonumber  \vspace{2mm}\\[8pt] 
\dfrac{dP}{dt} &=& \lambda_P E -   \sigma P,  \label{model:1 reduced} \vspace{2mm}\\[8pt]  
\dfrac{dI}{dt} &=& \lambda_I E +  \sigma P  - ( \psi_I+ \gamma_{I} ) I,   \nonumber \vspace{2mm}  \\[8pt]  
\dfrac{d H}{dt} &=& \psi_I I -(\psi_H +\gamma_H ) H, \nonumber \vspace{2mm}\\[8pt] 
\dfrac{d C}{dt} &=& \psi_H H  -(\gamma_C + \delta_C ) C, \nonumber \vspace{2mm}\\[8pt]
\dfrac{dR}{dt} &=& \gamma_A A + \gamma_{I} I + \gamma_{H} H + \gamma_C C.  \nonumber
\end{eqnarray} 
The basic reproduction number for the above system (\ref{model:1 reduced}) is given by
\begin{align}
\scalemath{0.91}{R_0 =\D\frac{1}{\left( \lambda_{{A}}+\lambda_{{P}}+
	\lambda_{{I}} \right)}\left[ {\D\frac {\beta_{{A}}\lambda_{{A}}}{  \gamma_{{A}}}}+{\D\frac {\beta_{{P}}\lambda_{{P}}}
	{  \sigma}}+{
	\D\frac {\beta_{{I}} \left( \lambda_{{I}}+\lambda_{{P}} \right) }{
		 \left( \psi
		_{{I}}+\gamma_{{I}} \right) }}+{\D\frac {\beta_{{H}}\psi_{{I}} \left( 
		\lambda_{{I}}+\lambda_{{P}} \right) }{ \left( \psi_{{I}}+\gamma_{{I}}
		\right) 
		\left( \psi_{{H}}+\gamma_{{H}} \right) }}\right] \label{R0 in reduced system}}
\end{align}
Note that the above expression of $R_0$ can be calculated from the expression of $R_C$ in (\ref{R_C formula}) by putting the control parameters, $\rho_S$ and $\varepsilon$, equal to zero.\\
\\
For comparison purposes between different control strategies, we also calculated the reproductive number when only one control of either isolation or quarantine is used, $R_{CJ}$ and $R_{CQ}$, and found to be
\begin{dmath}
	$$ R_{CJ}=\D\frac{1}{\left( \lambda_{{A}}+\lambda_{{P}}+\lambda_{{I}} \right)}\left[ {\D\frac {\beta_{{A}}\lambda_{{A}}}{   
			\gamma_{{A}}}} + \D\frac{\beta_P \lambda_P}{\sigma} + \D\frac{\beta_I \left( \lambda_I + \lambda_P\right)}{\left( \varepsilon+\psi_{{I}}+\gamma_{{I}} \right)} + \D\frac{\varepsilon\,\varepsilon_{
			{J}}\beta_{{J}} \left( \lambda_{{I}}+\lambda_{{P}}
		\right) }{\left( \varepsilon+\psi_{{I}}+ 
		\gamma_{{I}} \right)  \left( \psi_{{J}}+\gamma_{{J}} \right)} +
	\D\frac{\beta_{{H}} \left( \lambda_{{I}}+\lambda_{{P}}
		\right)  \left( \psi_{{I}} \left( \psi_{{J}}+\gamma_{{J}} \right) +
		\psi_{{J}}\varepsilon \right)}{\left( 
		\varepsilon+\psi_{{I}}+\gamma_{{I}} \right)  \left( \psi_{{J}}+\gamma_{{J
		}} \right)  \left( \psi_{{H}}+\gamma_{{H}} \right)} \right] $$ \label{R_CJ formula}
\end{dmath}
and
\begin{dmath}
	 R_{CQ}=\D\frac{\rho_Q}{\left(\rho_{{Q}}+\rho_{{S}} \right)\left( \lambda_{{A}}+\lambda_{{P}}+\lambda_{{I}} \right)}\left[ {\D\frac {\beta_{{A}}\lambda_{{A}}}{   
			\gamma_{{A}}}} + \D\frac{\beta_P \lambda_P}{\sigma} + \D\frac{\beta_I \left( \lambda_I + \lambda_P\right)}{\left( \psi_{{I}}+\gamma_{{I}} \right)}  +
	\D\frac{\beta_{{H}} \left( \lambda_{{I}}+\lambda_{{P}}
		\right)   \psi_{{I}} 
		}{\left( 
		\psi_{{I}}+\gamma_{{I}} \right)   \left( \psi_{{H}}+\gamma_{{H}} \right)} \right]  \label{R_CQ formula}
\end{dmath}
Formula \ref{R_CJ formula} can also be derived by setting the quarantine parameter $\rho_S=0$ in the $R_C$ expression. Similarly, one can get formula \ref{R_CQ formula} by putting the isolation parameter $\varepsilon=0$ in $R_C$ in (\ref{R_C formula}).\\ 
We shall use these expressions of $R_0$, $R_{CJ}$,  $R_{CQ}$  and $R_{C}$ in latter sections to compare the situation with and without control interventions.
\newpage
\section{ Fitting of Model Parameters and Numerical Results}
\begin{figure}[H]
	\centering
	\includegraphics[scale=0.75]{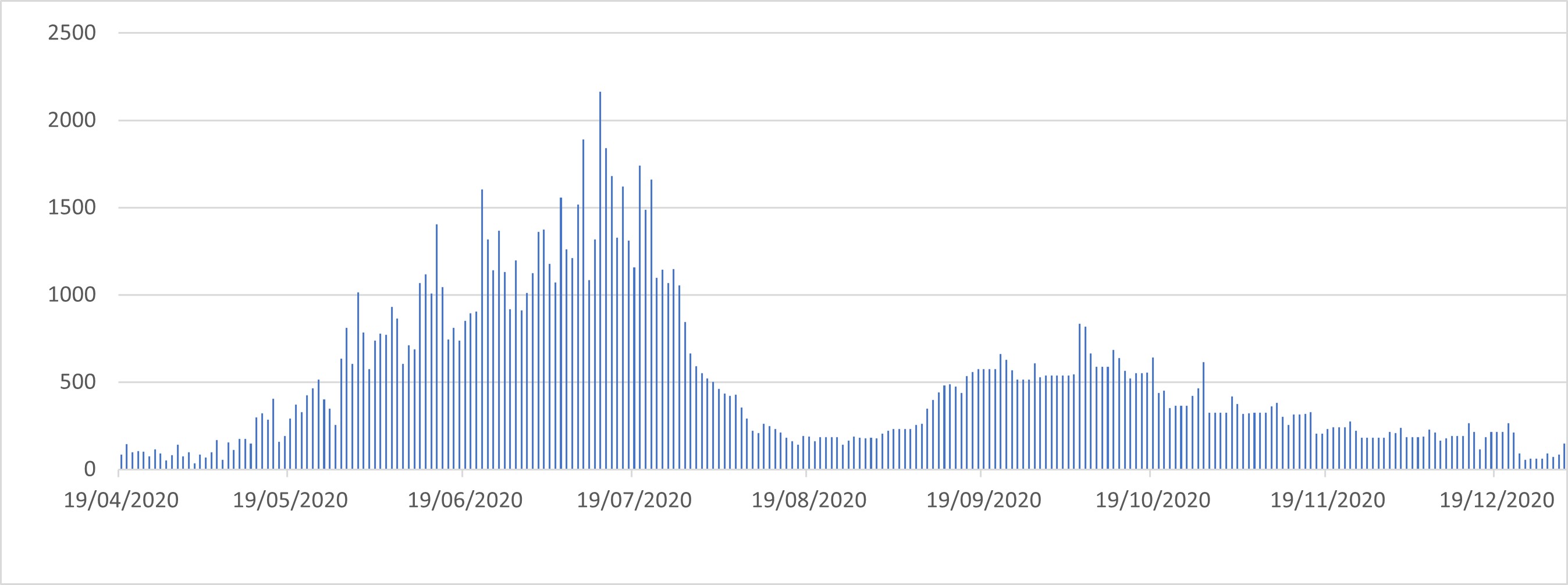}
	\caption{Oman daily cases from April 19 to December 21, 2020, with averaging data of unpublished days. }
	\label{Fig: OmanData2020}
\end{figure}

\begin{figure}[H]
	\centering
	\includegraphics[scale=0.9]{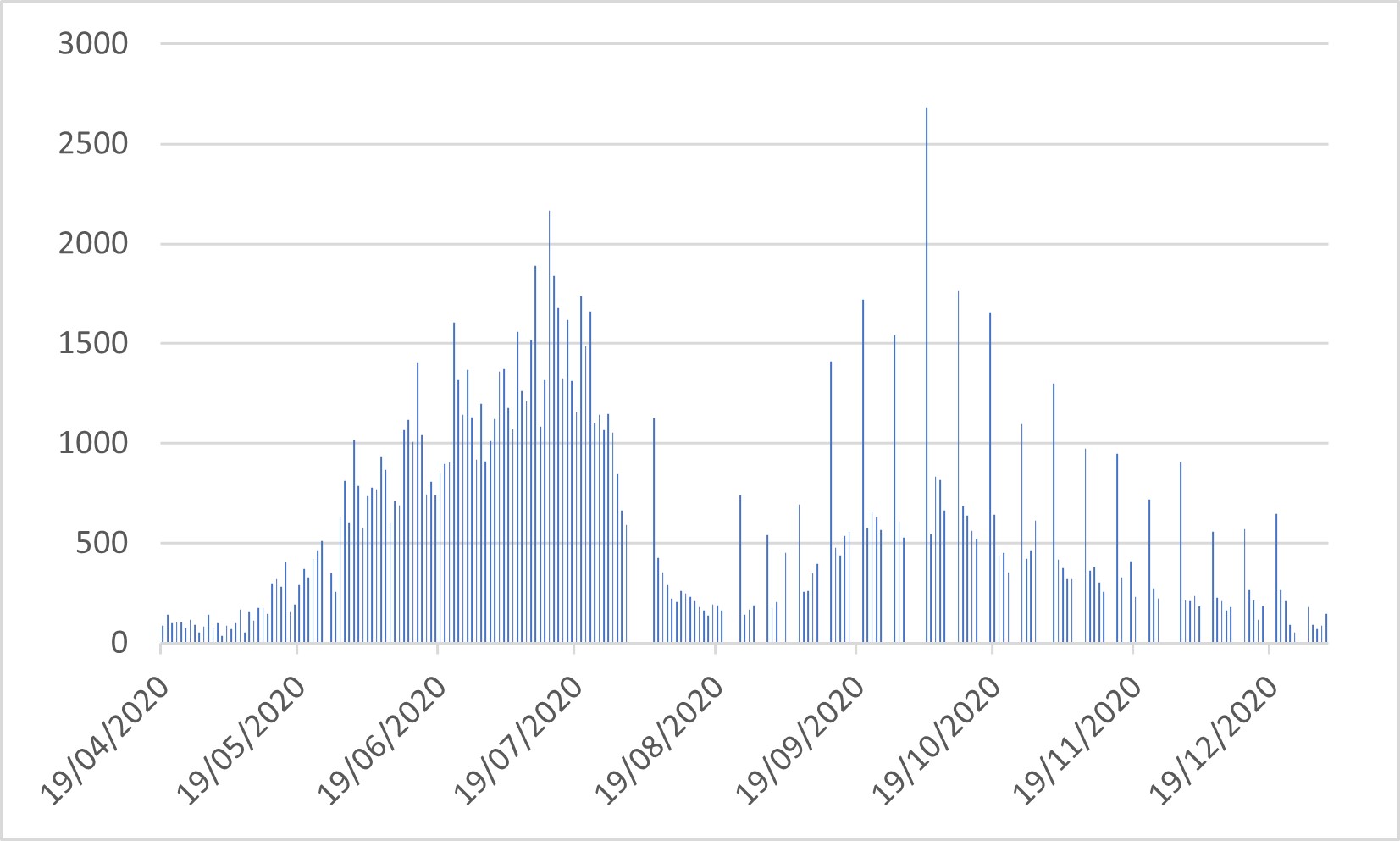}
	\caption{Oman daily cases from April 19 to December 21, 2020, without averaging data of unpublished days. }
	\label{Fig: OmanData2020_Noweekends}
\end{figure}

In this section, we use Oman data for estimating the 22 model parameters. Two data sets are used, representing part of the first and second waves of the pandemic that hit Oman. On February 24, 2020, the Oman Ministry of Health registered the first two COVID-19 cases for Omani women coming from Iran \cite{WinNT2}. Since then, the ministry of health started publishing daily reports about new and cumulative cases, deaths and recoveries until July 30, 2020. After that, no reports were published during public holidays and weekends. Starting from August 11, 2020, numbers of cumulative cases, deaths and recoveries were announced, but not the daily new ones. No data about hospital and ICU admission cases were published before June  4, 2020. This variation in publishing data made the fitting process complicated until a good fit was reached. As a result, comparing the dynamics of the two waves based on the estimated parameters was challenging.\\
The chart in Fig. \ref{Fig: OmanData2020} depicts the daily reported new cases from April 19 to December 21 , 2020. The method of averaging was used for the unreported data when producing this chart. The other chart in Fig. \ref{Fig: OmanData2020_Noweekends} is produced using reported data only. The two waves are apparent in both charts.
\subsection{Data Fitting and Parameters' Estimation}
\begin{figure}[H]
	\subfigure{\includegraphics[scale=0.27]{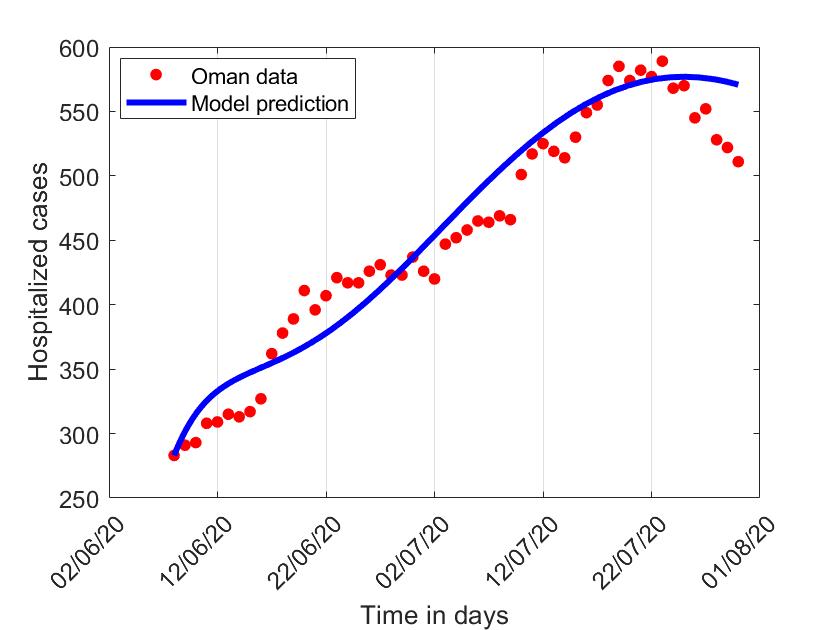}}
	\hfill
	\subfigure{\includegraphics[scale=0.27]{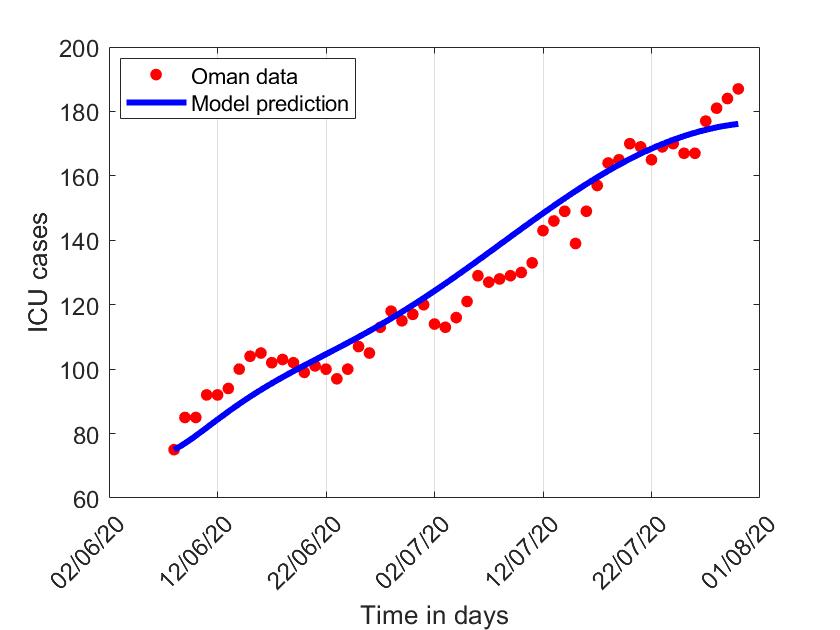}}
	\hfill
	\subfigure{\includegraphics[scale=0.27]{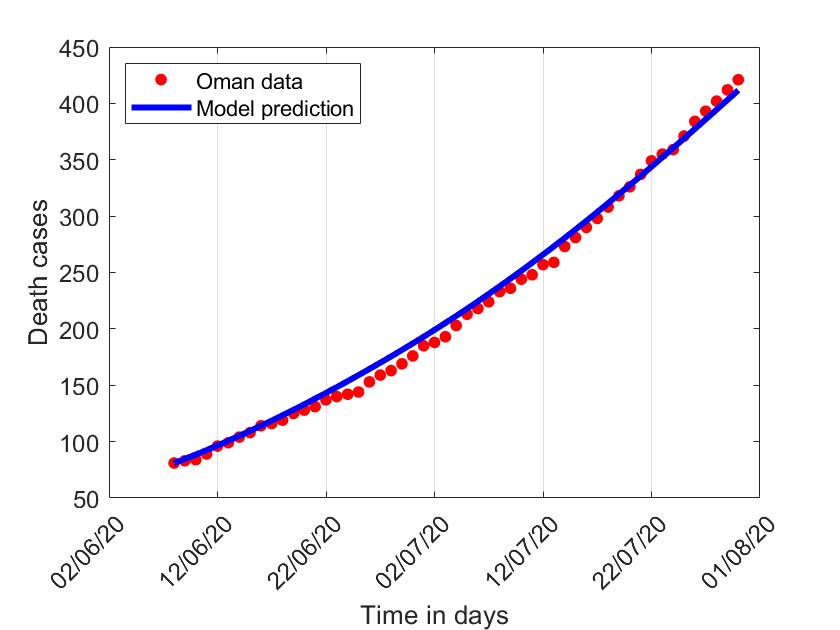}}
	\hfill
	\subfigure{\includegraphics[scale=0.27]{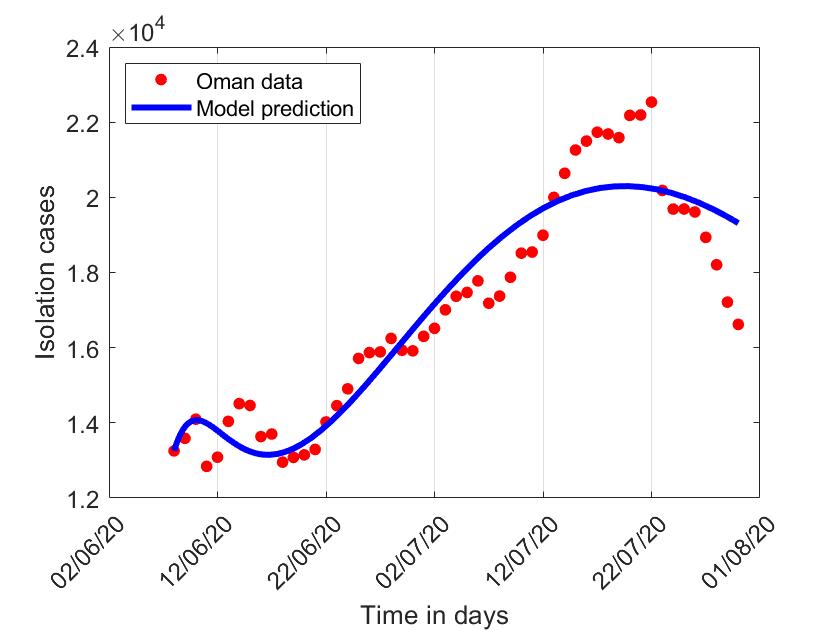}}
	\caption{Model fitting vs reported cases in Oman for the first wave.} \label{Fig: fitting w1}
\end{figure}

\begin{figure}[H]
	\subfigure{\includegraphics[scale=0.27]{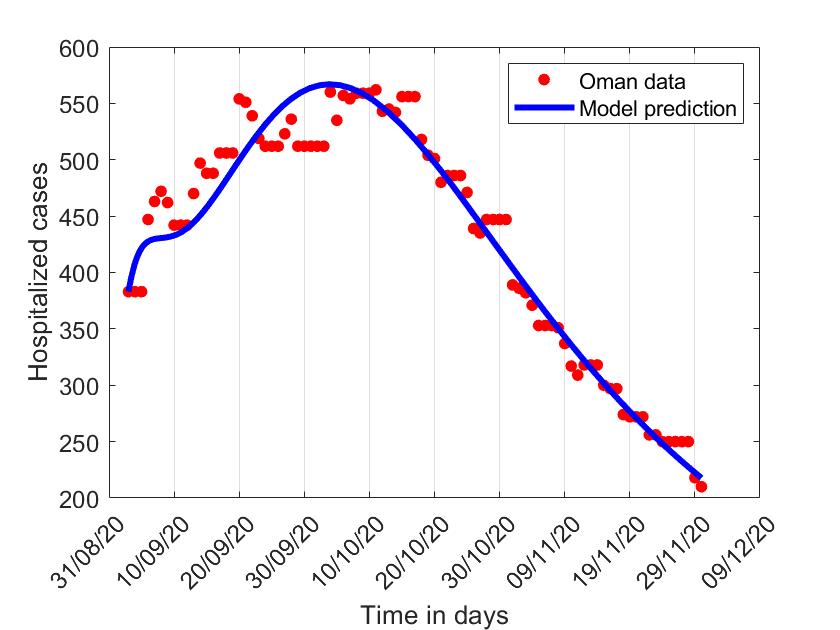}}
	\hfill
	\subfigure{\includegraphics[scale=0.27]{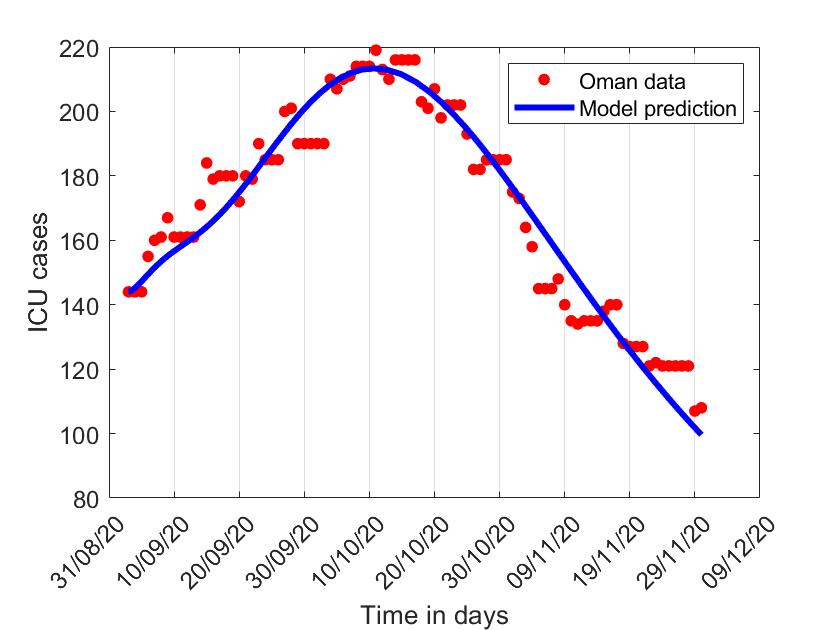}}
	\hfill
	\subfigure{\includegraphics[scale=0.27]{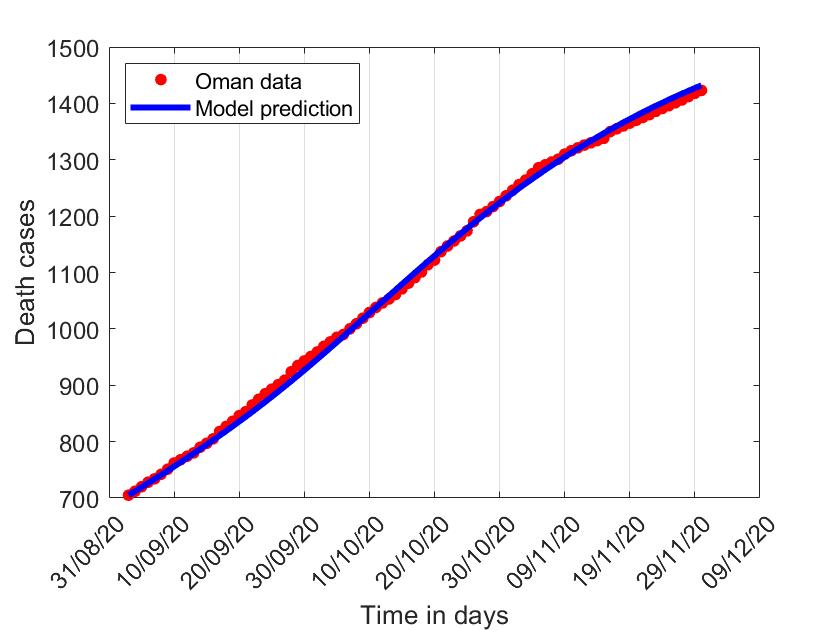}}
	\hfill
	\subfigure{\includegraphics[scale=0.27]{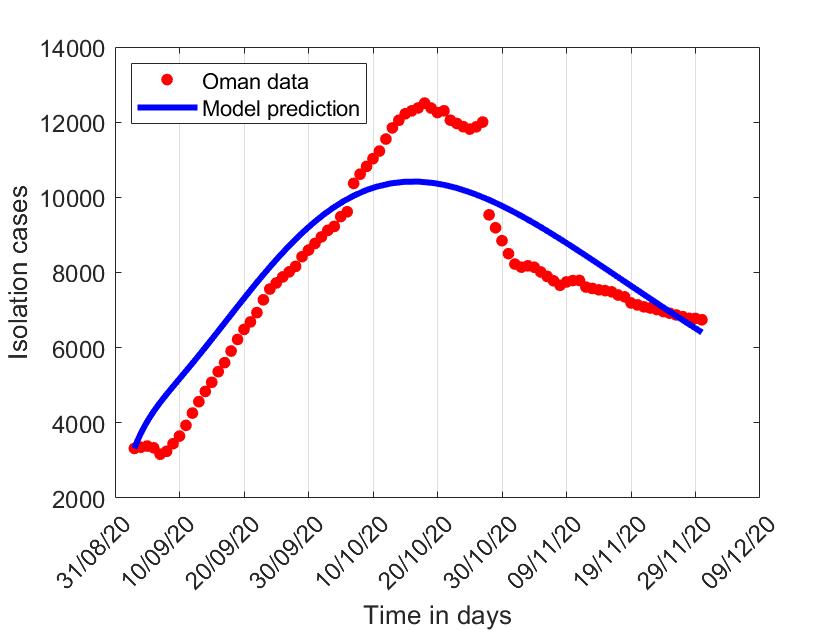}}
	\caption{Model fitting vs reported cases in Oman for the second wave.} \label{Fig: fitting w2}
\end{figure}

In this sub-section, we estimate the model parameters by fitting Oman data to the model equations. Oman data are collected from reports announced on the government's official twitter accounts of the Ministry of Health \cite{WinNT3} and the  efforts of countering COVID-19 \cite{WinNT4}.  Specifically, we used the data of hospitalization, intensive care unit, death and isolation. For isolation data, it is not officially published. However, assuming that all reported new cases are put in isolation by regulations, we computed the isolated cases from the cumulative, recovered, hospitalization, ICU and death cases so that: isolation = cumulative - (recovered + hospitalization+ICU+death).\\
\\
\noindent The graphs in Fig. \ref{Fig: fitting w1} compare Oman reported data with the model prediction for the period June 8 to July 30, 2020, within the first wave of COVID-19 in Oman. 
\noindent The graphs in Fig. \ref{Fig: fitting w2} display Oman COVID-19 data together with the model prediction for the period September 3, 2020, to November 30, 2020, within the second wave of the pandemic in Oman.
\noindent Best data fitting was attempted by including more than a class of data of COVID-19 clinical states in the non-linear least square curve fitting approach. Also, we avoided including that data when cases of more than three successive days were not reported. Moreover, when using the Matlab optimization function fminsearchbnd, we provided the initial value guess for the 22 fitted parameters from previous studies. We also set the upper and lower bounds for each parameter by taking into consideration reasonable rates together with possible population portions.\\
The obtained values for the estimated parameters with the best fit are listed in Table \ref{tablel: fitting}.
\begin{table}[H]
	\caption{Fitted values for parameters in the model (\ref{model:1}).}
	\label{tablel: fitting}
	\centering
	\begin{tabular}{|c|c|c|}
		\hline
		Parameter  & Fitted value & Fitted value \\
		& wave 1 & wave 2\\ \hline
		$\beta_A$   &0.0015 & 0.0014 \\ 
		$\beta_P$   &0.2328 &0.4394 \\ 
		$\beta_I$   & 0.8967 & 0.8127  \\
		$\beta_J$  &0.013 & 0.2476 \\ 
		$\beta_H$ &  0.0018 & 0.0035 \\ 
		$\varepsilon_J$ & 0.6656& 0.629\\
		$\rho_S$ & 0.0233 & 0.0272 \\
		$\rho_Q$ &0.0226 & 0.0167  \\
		$\lambda_A$  & 0.0743 & 0.1787 \\
		$\lambda_P$ & 0.0659 & 0.0534\\
		$\lambda_I$  & 0.0309 & 0.0348\\
		$\sigma$ & 0.6052 & 0.2786 \\
		$\varepsilon$  & 0.1695 & 0.1343\\
		$\psi_I$ & 0.0015 & 0.0207 \\
		$\psi_J$  & 0.0025& 0.0003 \\
		$\psi_H$ & 0.0389 & 0.0534 \\
		$\gamma_A$  &0.0968 &0.1998  \\
		$\gamma_I$ &0.1535&0.1267 \\
		$\gamma_J$ & 0.1171&0.0436 \\
		$\gamma_H$  & 0.083 & 0.0993 \\
		$\gamma_C$  & 0.0748 &0.0898 \\
		$\delta_C$   & 0.049&0.0484 \\
		\hline 
	\end{tabular}
\end{table}
\noindent One can observe from Table \ref{tablel: fitting}  that the transmission rate from individuals with clinical symptoms, $\beta_I$, is the highest and thus accounts for the majority of the infections in both first and second waves. The transmission from the isolated group, $\varepsilon_J \beta_J$, is higher in the second wave compared to the first one. This can be interpreted, maybe in light of 
regulation in the second wave varied from that in the first one. For Oman, strict institutional isolation was implemented in the first stage of the pandemic, and a positive tested individual had to be isolated at a  place provided by the government, mostly at hotels, for those who did not have a suitable place in their accommodation. Those who had a suitable private place to isolate themselves at home were tracked by electrical bracelets. On the opposite side, when the regulations were eased and relaxed  months later, fewer people were in institutional isolation, and the majority were isolated at home. It is by common sense that in home isolation, more contact with others is expected and hence more transmission results. This justifies the higher transmission rate from the isolation class, $\beta_J$, in the second wave compared with the first.\\
To justify the above reasoning, we collected available data about the number of institutional isolation centres and isolated people. Only data shown in Table \ref{table: Institutional Isolation} were officially announced.
\begin{table}[H]
	\caption{Number of active institutional isolation centres with isolated people.  }
	\label{table: Institutional Isolation}
	\centering
	\begin{tabular}{|l|c|c|c|}
		\hline
		date & isolation centres& people in isolation& source \\
		\hline
		14 July, 2020&31&1525&\cite{WinNT14Jul}\\
		\hline
		18 July, 2020&26&1473&\cite{WinNT18Jul}\\
		\hline
		22 July, 2020&29&756&\cite{WinNT22Jul}\\
		\hline
		16 August, 2020&4&17&\cite{WinNT16Aug}\\
		\hline
		1 September, 2020&1&6&\cite{WinNT1Sep}\\
		\hline
	\end{tabular}
\end{table}
\noindent It is obvious from the above table how the number of active institutional centres and people at them were more during the time of the first wave compared to the second wave.\\
\noindent This finding that home isolation is not as powerful as institutional isolation supports the study by BL Dickens et al. \cite{dickens2020institutional} in that institutional, not home-based, isolation could contain the COVID-19 outbreak.\\
However, variation in the effectiveness of isolation, whether institutionally or at home, between the two waves can also be due to other causes influencing the isolation process. The approach in Wuhan and the nearby cities in Hubei Province was examined \cite{niu2020deciphering} and found that initial outbreak sizes with  early detection and response were among the key determinants for the success of isolation. It is also found that transmission by people with no or mild symptoms can dampen the power of the isolation strategy because of the reduced likelihood of isolating all cases and tracing all contacts. Another significant challenge for the completeness in case isolation was that nucleic acid testing, the primary tool for case identification, had a variable rate of false-negative 
results, so even symptomatic cases could be set free,  
thus weakening the feasibility of controlling COVID-19 
outbreaks by isolation of cases and contacts.\\
\\
\noindent Based on the estimated values of the 22 parameters in Table \ref{tablel: fitting} and using formula \ref{R_C formula} for approximating the control reproduction number, $R_C$, we computed the transmission contribution for each route of $R_C$ for both waves, as shown in Table \ref{table: transmission routes}.
\begin{table}[H]
	\caption{Transmission routes' contribution to $R_C$.   }
	\label{table: transmission routes}
	\centering
	\begin{tabular}{|c|c|c|c|c|c|c|}
		\hline
		&$R_A$ & $R_P$& $R_I$& $R_J$ & $R_H$&$R_C$\\
		\hline
		Wave 1&0.00331&0.07295&0.76976&0.00613&0.00006&0.85221\\
		\hline
		Wave 2&0.00178&0.12004&0.36267&0.15816&0.00022&0.64287\\
		\hline
	\end{tabular}
\end{table}
\noindent For model (\ref{model:1}), it is evident from Table \ref{table: transmission routes}  that the contribution from the symptomatic individuals, $R_I$, is the highest in both waves. This finding is reasonable as infectees in the class ($I$) are not known or identified and, accordingly, are not isolated by authorities' orders. Hence, they can move freely and spread the infection to many others.\\
During the first wave, the contribution from the presymptomatic individuals, $R_P$, is higher than that of the isolated group, $R_J$. This result is also acceptable as infected people during the early stage of the pandemic were more cautious and tended to isolate themselves to avoid infecting other people. Conversely, during the second wave, the contribution from the isolated class, $R_J$, is slightly higher than the contribution from the presymptomatic class, $R_P$, and the reason is, as mentioned earlier, regarding human behaviour.\\
Contributions from asymptomatic, $R_A$, and hospitalized, $R_H$, classes are small during both waves. This is because the viral load in asymptomatic individuals is low, and as a result, the chance of successful transmission becomes low. Also, hospitalized patients are considered in strict isolation and the transmission, if it occurs, is only to health workers.\\
\\
To study the scenario if less or no control were implemented, we calculated the reproduction number and transmission from each route, assuming there was no control at all (Table \ref{table: transmission routes R_0}), isolation was the only control measure (Table \ref{table: transmission routes R_CJ}), and quarantine was the only control adopted (Table \ref{table: transmission routes R_CQ}). The values in Tables \ref{table: transmission routes R_0}, \ref{table: transmission routes R_CJ} and \ref{table: transmission routes R_CQ} are computed using formulas \ref{R0 in reduced system}, \ref{R_CJ formula}
and \ref{R_CQ formula}, respectively.
\begin{table}[H]
	\caption{Transmission routes' contribution to $R_{0}$ (without  control).  }
	\label{table: transmission routes R_0}
	\centering
	\begin{tabular}{|c|c|c|c|c|c|c|}
		\hline
		&$R_A$ & $R_P$& $R_I$& $R_J$ & $R_H$&$R_0$\\
		\hline
		Wave 1&0.00673&0.14816&3.27296&0&0.00008&3.42793\\
		\hline
		Wave 2&0.00469&0.31555&1.82202&0&0.00106&2.14332\\
		\hline
	\end{tabular}
\end{table}
\begin{table}[H]
	\caption{Transmission routes' contribution to $R_{CJ}$ (with isolation only).  }
	\label{table: transmission routes R_CJ}
	\centering
	\begin{tabular}{|c|c|c|c|c|c|c|}
		\hline
		&$R_A$ & $R_P$& $R_I$& $R_J$ & $R_H$&$R_{CJ}$\\
		\hline
		Wave 1&0.00673&0.14816&1.56336&0.01246&0.00013&1.73083\\
		\hline
		Wave 2&0.00469&0.31555&0.95337&0.41576&0.00058&1.68996\\
		\hline
	\end{tabular}
\end{table}
\begin{table}[H]
	\caption{Transmission routes' contribution to $R_{CQ}$ (with quarantine only).  }
	\label{table: transmission routes R_CQ}
	\centering
	\begin{tabular}{|c|c|c|c|c|c|c|}
		\hline
		&$R_A$ & $R_P$& $R_I$& $R_J$ & $R_H$&$R_{CQ}$\\
		\hline
		Wave 1&0.00331&0.07294&1.61152&0&0.00003&1.68782\\
		\hline
		Wave 2&0.00178&0.12004&0.69311&0&0.0004&0.81534\\
		\hline
	\end{tabular}
\end{table}

\noindent It is clear from Table \ref{table: transmission routes R_0} how the situation could be worse if no control measures were imposed with the basic reproduction number $R_0 \approx 3.42793$ and $R_0 \approx 2.14332$ in the first and second wave, respectively. If isolation alone was implemented,  the reproduction number might have decreased to $R_{CJ} \approx 1.73083$ and $R_{CJ} \approx 1.68996$ for the first and second wave, respectively. Hence, isolation alone as a mitigation strategy would not contain the spread. If quarantine only were to be introduced, then the estimated reproduction number would reduce to  $R_{CQ} \approx 1.68782$ and  $R_{CQ} \approx 0.81534$ for the first and second wave, respectively. Therefore, quarantine alone would be sufficient to eliminate the disease during the second wave but not the first. According to Oman data and since both isolation and quarantine were implemented, the estimated reproduction number $R_C \approx 0.85221$ and $R_C \approx 0.64287$ in the first and second wave, respectively, which led to the temporary elimination of the disease after each wave.\\
One can also conclude from the tables above that introducing isolation alone did not affect the transmission from asymptomatic and presymptomatic classes, as it is clear from the values of $R_A$ and $R_P$ in Tables \ref{table: transmission routes R_0} and \ref{table: transmission routes R_CJ}. However, isolation helped decreasing the transmission from the symptomatic class by almost half its contribution without any control, as shown by the values of $R_I$ in Tables \ref{table: transmission routes R_0} and \ref{table: transmission routes R_CJ}. \\
In contrast, using quarantine alone as a control measure could reduce the transmission from the asymptomatic and presymptomatic classes by more than two times, as given by the values of $R_A$ and $R_P$ in Tables  \ref{table: transmission routes R_0} and \ref{table: transmission routes R_CQ}. 
\subsection{Sensitivity Analysis}
\begin{figure}[H]
	\subfigure[First wave]{
		\includegraphics[scale=0.246]{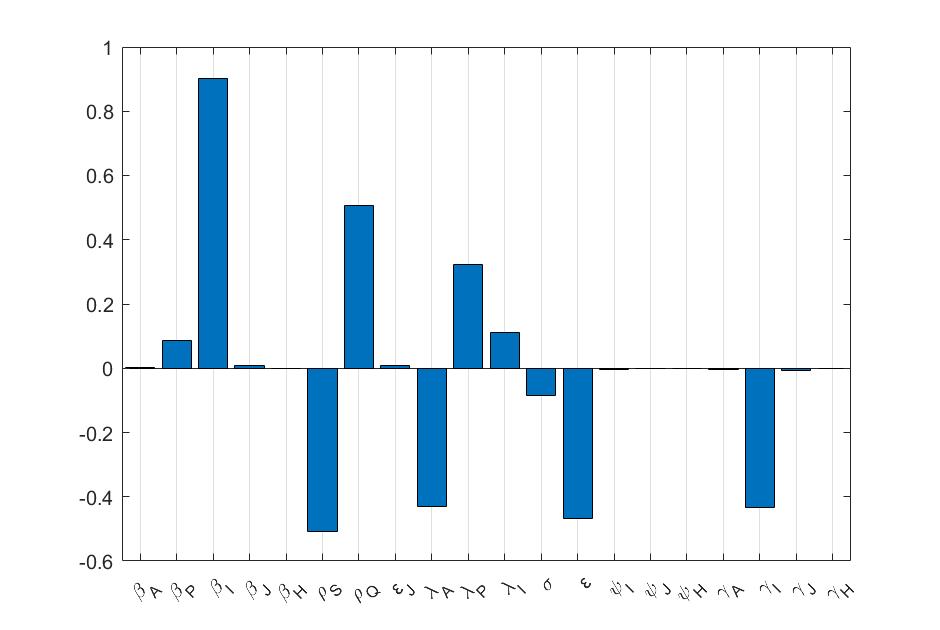}}
	\subfigure[Second wave ]{
		\includegraphics[scale=0.246]{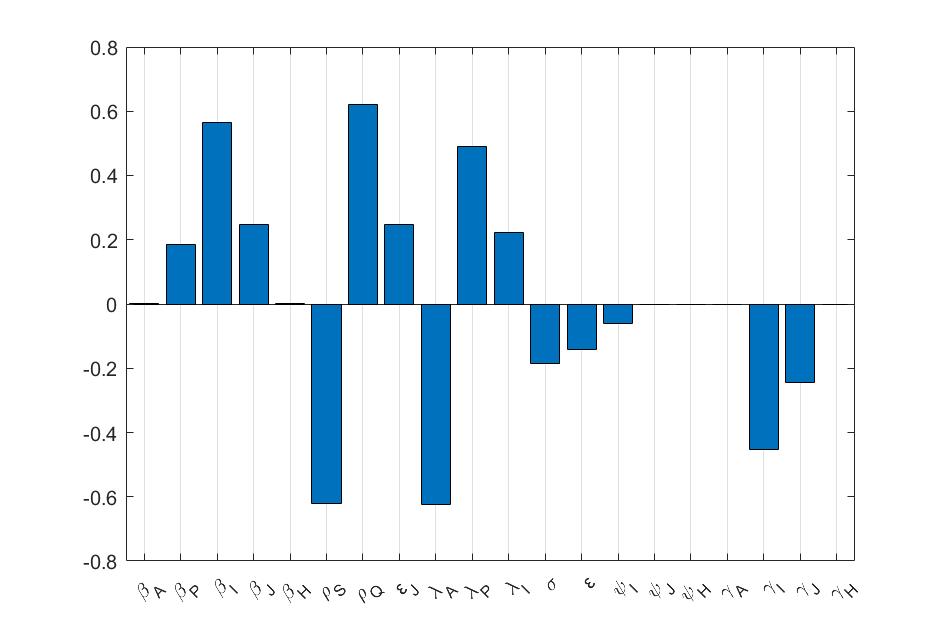}}
	\caption{Sensitivity chart for the first and second waves.} \label{Fig: sensitivity w2}
\end{figure}

This section uses sensitivity analysis to study the influence of different models' parameters on the spread of COVID-19. The sensitivity index \cite{van2017reproduction} of $R_C$ with respect to a parameter $\varphi$ is $\D\frac{\partial R_C}{\partial \varphi }$. The
normalized sensitivity index (also known as the elasticity index) measures the relative change of $R_C$ with respect to $\varphi$, denoted by $\Upsilon_{\varphi}^{R_{C}}$, and is defined as $$\Upsilon_{\varphi}^{R_{C}}= \D\frac{\partial R_C}{\partial \varphi } \D\frac{\varphi}{R_C}.$$ 
Using the expression of $R_C$ in (\ref{R_C formula}), we computed the elasticity index for each parameter and evaluated at the two sets of estimated parameters in Table \ref{tablel: fitting}. The resulted elasticity index values for each parameter for both waves are listed in Table \ref{table: sensitivity}. A chart view of the elasticity indices is also provided in Fig. \ref{Fig: sensitivity w2}. The positive/negative sign of the elasticity indicates that $R_C$ increases/decreases with the increase of the parameter, whereas the magnitude determines the relative impact of the parameter. These indices are essential in deciding which control strategies to implement by indicating the most influencing parameters to target.
\noindent In both waves, it is clear that $R_C$ is more positively sensitive to small changes in the transmission rate of symptomatic individuals $\beta_I$, rate of leaving quarantine, $\rho_Q$, and the rate at which exposed become pre-symptomatically and symptomatically infectious, $\lambda_P$ and $\lambda_I$, respectively. For the second wave, $R_C$ is more positively sensitive to parameters related to transmission from the isolated group, $\varepsilon_J$ and $\beta_J$, compared to the first wave. \\
$R_C$ is more negatively sensitive to the quarantine rate $\rho_S$, the rate at which exposed individuals become asymptomatically infectious $\lambda_A$ and the recovery rate of symptomatic individuals $\gamma_{I}$. During the first wave, $R_C$ is more negatively sensitive to the isolation rate of symptomatic infectees $\varepsilon$ compared to the second wave. During the second wave, $R_C$ is more negatively sensitive to the recovery rate of isolated people $\gamma_J$ compared to the first wave.
\begin{table}[H]
	\caption{Sensitivity analysis of model (\ref{model:1}). }
	\label{table: sensitivity}
	\centering
	\begin{tabular}{|c|l|l|}
		\hline
		Parameter $ ( \varphi )$ &  $\Upsilon_{\varphi}^{R_{c}}$ (wave1) &   $\Upsilon_{\varphi}^{R_{c}}$ (wave2)\\
		\hline
		$\beta_A$& 0.003888& 0.002776  \\
		& &  \\
		$\beta_P$&0.085598 & 0.186722  \\
		& &  \\
		$\beta_I$&0.903243 &0.564141  \\
		& &  \\
		$\beta_J$&0.007196 &0.246017  \\
		& &  \\
		$\beta_H$&0.000075 &0.000344 \\
		& &  \\
		$\rho_S $&-0.507625 &-0.61959 \\
		& &  \\
		$\rho_Q$&0.507625 &0.61959 \\
		& &  \\
		$ \varepsilon_J $&0.007196  & 0.246017 \\
		& & \\
		$\lambda_A$&-0.429057 & -0.624574 \\
		& &  \\
		$\lambda_P$&0.321465 & 0.489966 \\
		& &  \\
		$\lambda_I$&0.111353& 0.221481 \\
		& &  \\
		$\sigma$ &-0.085598 & -0.186722 \\
		& &  \\
		$\varepsilon$ &-0.468351 & -0.140373  \\
		& &  \\
		$\psi_I$ &-0.004187 & -0.059228  \\
		& & \\
		$\psi_J$& -0.000099 & -0.001667 \\
		& &  \\
		$\psi_H$&-0.000024 & -0.00012 \\
		& &  \\
		$\gamma_A$ &-0.003888 & -0.002776  \\
		& &  \\
		$\gamma_I $ &-0.434466& -0.451412  \\
		& & \\
		$\gamma_J$ &-0.007097 & -0.244351 \\
		& &  \\
		$\gamma_H$	& -0.000051 & -0.000224  \\
			\hline
	\end{tabular}
\end{table}
\newpage
\subsection{Numerical Simulation}
This section presents some numerical simulations of the model(\ref{model:1}) using the fitted values of parameters in Table \ref{tablel: fitting}. 
The simulations assess the impact of quarantine and isolation on the number of hospitalized cases. We observed the same dynamics when using ICU cases to study the impact of quarantine and isolation. Therefore, we provided only graphs that represent the effects on hospitalization. 
In each graph, the black curve is the baseline of the estimation, whereas the others are used for comparing the effects and predicting the scenario when the estimated baseline parameter is varied. For reasonable comparison, we varied most of the baseline parameters by a factor of 2. 
\subsubsection{Effects of Quarantine}
\begin{figure}[H]
	\subfigure[Effect of $\rho_S$ on the first wave]{\includegraphics[scale=0.25]{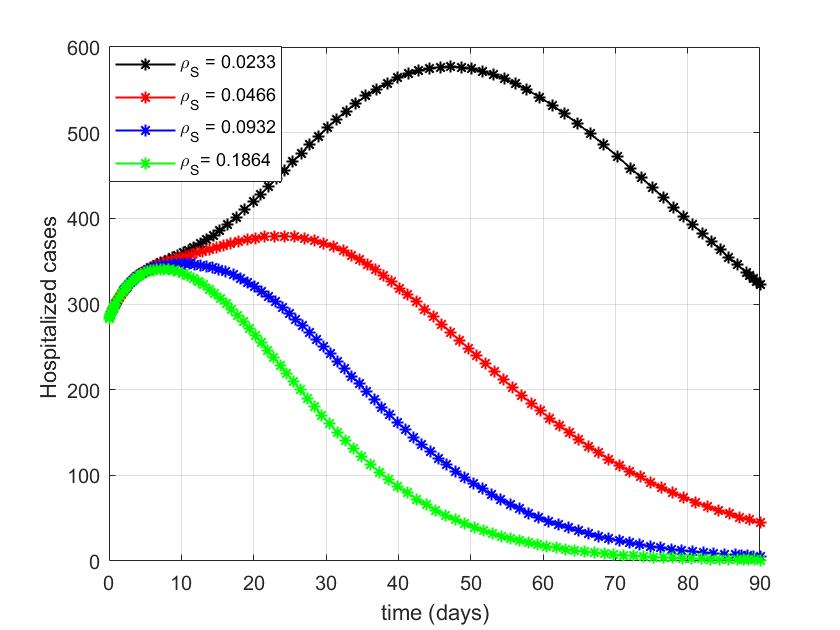}}
	\hfill
	\subfigure[Effect of $\rho_S$ on the second wave]{\includegraphics[scale=0.25]{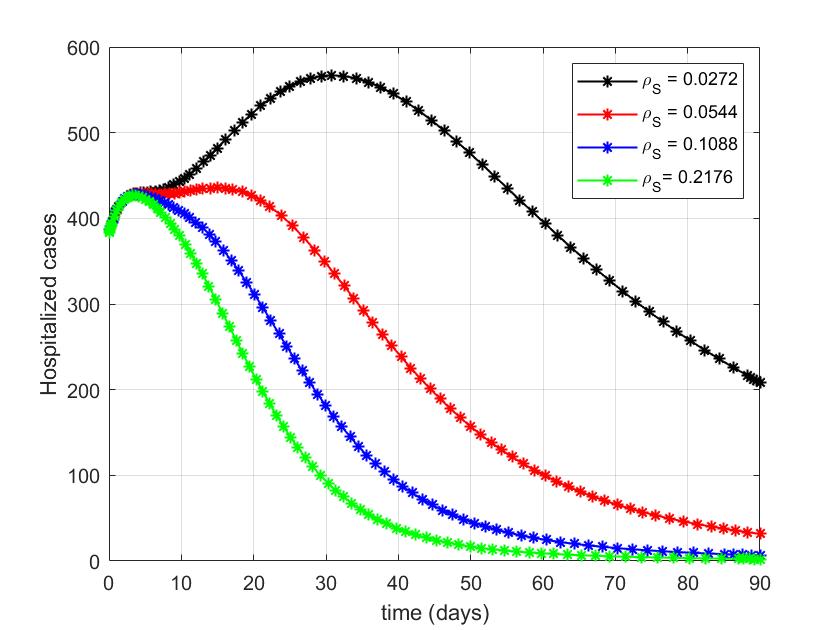}}
	\hfill
	\caption{Effects of quarantine rates on the number of hospitalization.} \label{fig: quarantine effects rho-s}
\end{figure}

The effects of quarantine are assessed by increasing or decreasing the quarantine rate of susceptible individuals $\rho_S$ and the rate of leaving quarantine $\rho_Q$. The effect on the number of hospitalization during the first and second waves is shown in Fig. \ref{fig: quarantine effects rho-s}a and \ref{fig: quarantine effects rho-s}b, respectively. In both waves, similar dynamics are obtained when increasing the baseline estimated quarantine rate by a factor of 2 at each simulation. The favourable consequence of increasing the quarantine rate is not only in reducing the number of hospitalization but also in bringing the lower peak ahead in time; consequently, the elimination of the disease takes a shorter time.
\noindent This conclusion is valid for our current epidemic model, where there is no recruitment in the susceptible population. In endemic models, where the population is recruited, it is expected that it may rise again after the elimination of the disease.
\begin{figure}[H]
	\subfigure[Effect of $\rho_Q$ on the first wave]{\includegraphics[scale=0.25]{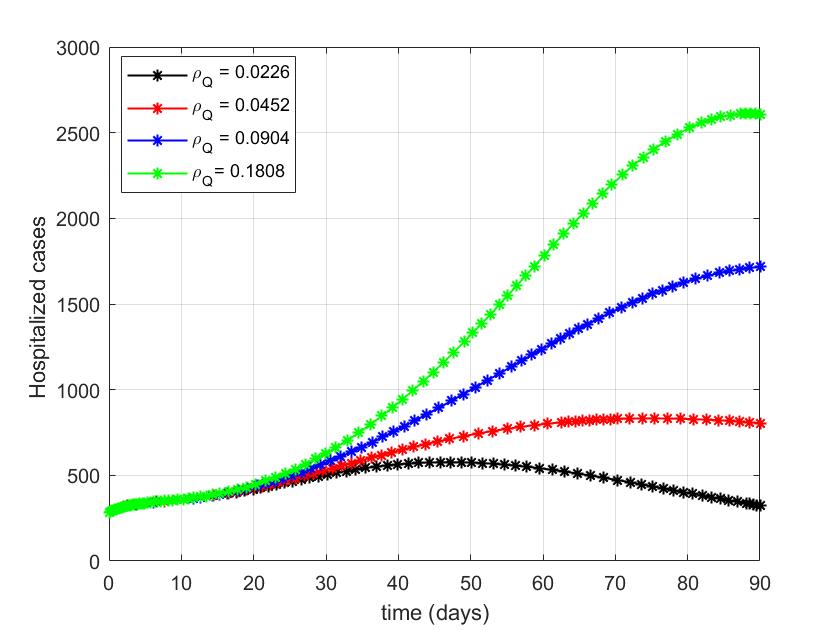}}
	\hfill
	\subfigure[Effect of $\rho_Q$ on the second wave]{\includegraphics[scale=0.25]{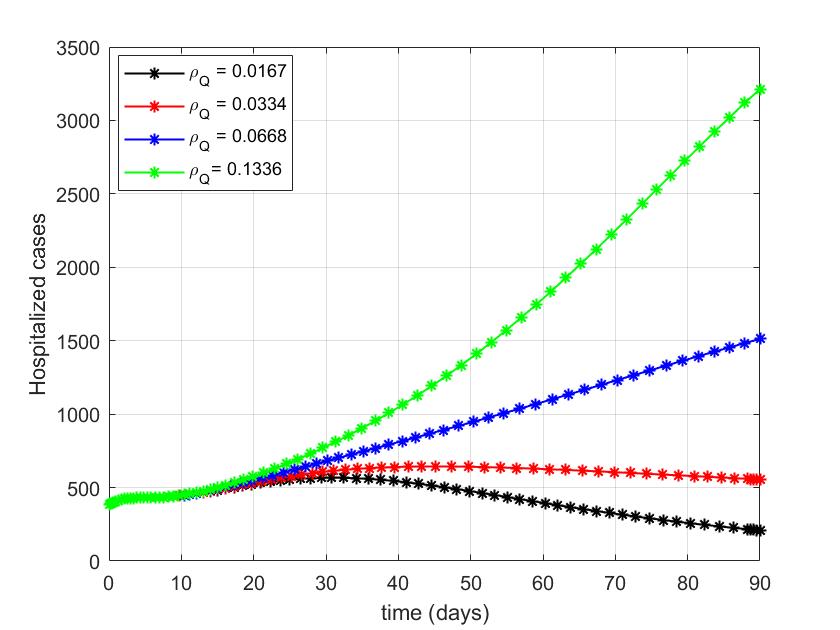}}
	\hfill
	\caption{Effects of rates of leaving quarantine on the number of hospitalization.} \label{fig: quarantine effects rho-q}
\end{figure}
\noindent The rate at which people leave their quarantine $(\rho_Q)$ has an opposite effect, as can be concluded from graphs of Fig. \ref{fig: quarantine effects rho-q}a and \ref{fig: quarantine effects rho-q}b. The faster the rate of leaving quarantine is the higher number of admission results. Thus, it is helpful to encourage people to keep staying at home when it is not necessary to leave it.
\subsubsection{Effects of Isolation}
The effects of isolation are examined by varying the baseline of estimated parameters related to isolation. Namely, the fraction of non-adherence to isolation parameter, $\varepsilon_J$,  and the rate at which symptomatic-infectious individuals are isolated, $\varepsilon$.
\begin{figure}[H]
	\subfigure[Effect of $\varepsilon_J$ on the first wave]{\includegraphics[scale=0.25]{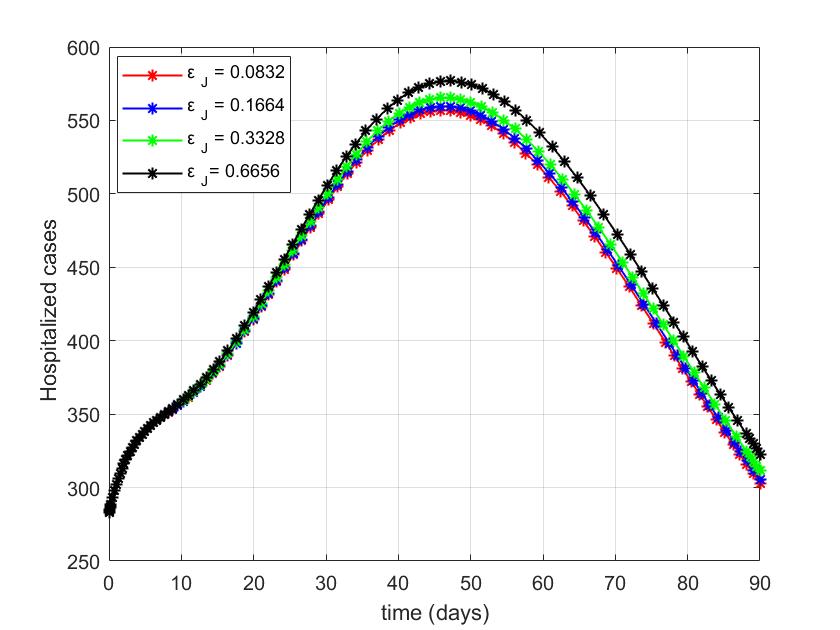}}
	\hfill
	\subfigure[Effect of $\varepsilon_J$ on the second wave]{\includegraphics[scale=0.25]{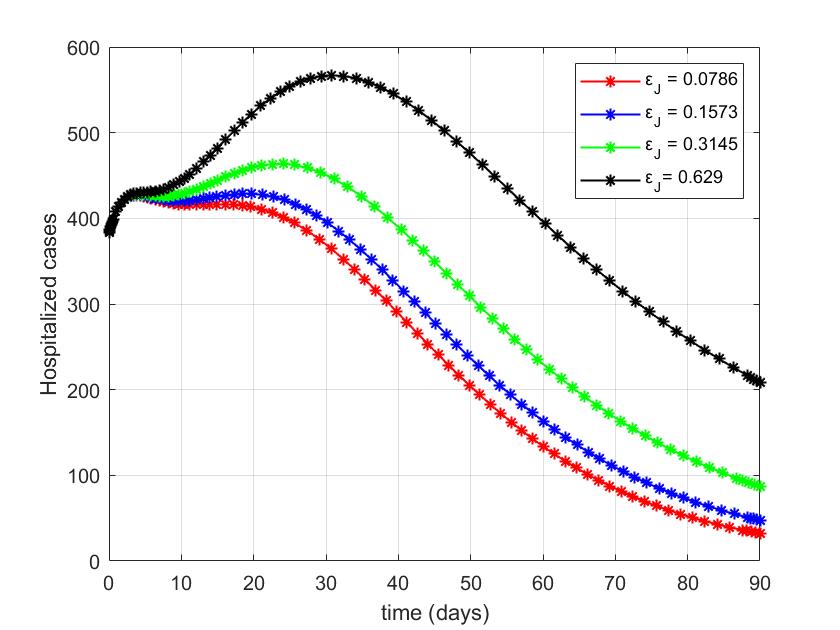}}
	\hfill
	\caption{Effects of adherence to isolation rules on the number of hospitalization.} \label{fig: isolation effects eps-j}
\end{figure}
\noindent Fig. \ref{fig: isolation effects eps-j}a and \ref{fig: isolation effects eps-j}b illustrate the effects of varying the fraction of non-adherence to isolation ($\varepsilon_J$) during the first and second waves, respectively. In both cases, $\varepsilon_J$ is varied by a factor of 2. However, the impact during the first wave is weak and much smaller compared with the considerable and robust impact during the second wave. Numerical simulation has nothing to say about this notable variation in the two waves as simulations agree with the theoretical analysis provided earlier. To demonstrate such a difference, although the fitted approximations of $\varepsilon_J$ of the two waves are close to each other, as given in Table \ref{tablel: fitting}, the estimated transmission rates from the isolated group ($\beta_J$) are far from each other. One justification could be that the COVID-19 variant during the second wave was much stronger than that of the first wave. Hence, even with the same level of isolation and adherence, the more potent variant causes more infections.
\begin{figure}[H]
	\subfigure[Effect of $\varepsilon$ on the first wave]{\includegraphics[scale=0.25]{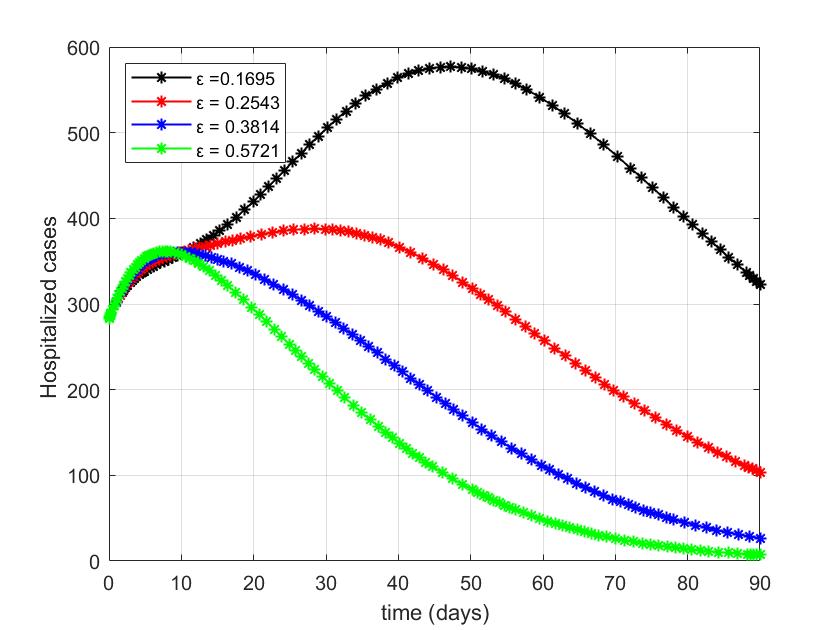}}
	\hfill
	\subfigure[Effect of $\varepsilon$ on the second wave ]{\includegraphics[scale=0.25]{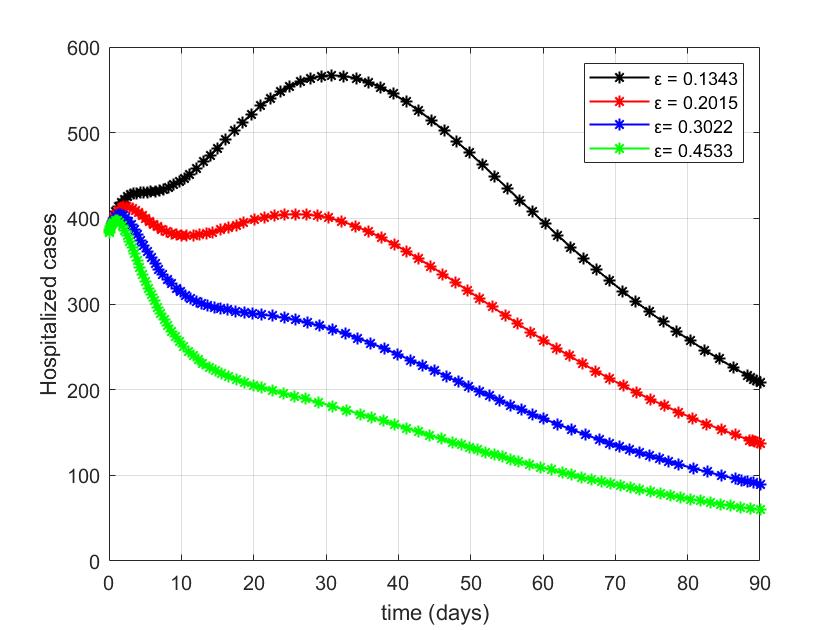}}
	\hfill
	\caption{Effects of isolation rates on the number of hospitalization. Parameters' values are varied by a factor of 1.5.} \label{fig: isolation effects eps}
\end{figure}
\noindent The rate at which infected individuals are identified and put in isolation, $\varepsilon$, has similar dynamics during the two waves, but with a stronger impact on the number of hospitalization during the first wave. Graphs of Fig. \ref{fig: isolation effects eps}a  and \ref{fig: isolation effects eps}b clarify how cases decreased significantly when the baseline estimated values of $\varepsilon$ are increased by a factor of 1.5. During the first wave, if the isolation rate is increased by a factor of 3.375 (from 0.1695 to 0.5721), the hospitalized cases drop from almost 600 to 100 on day 50. On the other hand, during the second wave, increasing the isolation rate by the same factor (from 0.1343 to 0.4533) results in decreasing the number of cases in hospitals from almost 600 to 200 on day 30. 
Therefore, it is crucial to identify the suspected infections by early testing to control the spread by isolating infectees.\\
\\
To sum up, the above numerical simulation supports the importance of both quarantine and isolation in reducing the number of infections. Certainly,  putting part of the susceptible population under quarantine by implementing partial or whole lockdown is very effective, as  deduced from Fig. \ref{fig: quarantine effects rho-s} and \ref{fig: quarantine effects rho-q}. However, this intervention has economic and social consequences for individuals and organizations. Therefore, it is more convenient to enhance isolation and minimize quarantine. We found that to reduce the quarantine rate ($\rho_S$) to  zero, the minimum isolation rate ($\varepsilon$) must be above $0.45832$ to eliminate the disease, which is the case when $R_C \le 1$.
\section{Conclusion}
In this work, we presented a mathematical model for assessing the impact of quarantine and isolation as NPI tools to control the spread of COVID-19. In particular, we analyzed their effects on the number of hospitalized and ICU cases in Oman. The positivity of solutions was proved. The proposed model has a continuum of disease-free equilibria, $\mathcal{E}_0$. It was shown that $\mathcal{E}_0$ is locally and globally asymptotically stable whenever the associated control reproduction number ($R_C$) is less than unity. For comparison purposes, we also derived the reproduction number assuming there was no control strategy or only one control method was imposed. The model has 22 parameters which were estimated by fitting model equations to Oman data. Two sets of data representing two waves of the pandemic were used to generate two sets of estimated parameters.\\
Based on the obtained fitted values of the parameters , the basic and control reproduction numbers, of different control-measures scenarios, were approximated. Moreover, the contribution to the reproduction number from transmission routes were calculated. \\ 
Sensitivity analysis was carried out to gain insight into vital parameters that considerably influence the reproduction number. We found that the transmission rate from the isolated group and, consequently, the contribution to the reproduction number have a more significant impact during the second wave than in the first one. We interpreted this variation in light of the human behaviour and the type of isolation (home vs institutional). We also found that the transmission from the symptomatic class was the highest, so identification of these individuals in earlier stages to isolate them is highly recommended.\\ Some numerical simulations were provided to evaluate the effects of quarantine and isolation on the number of hospitalized cases. These simulations support the find that using quarantine and isolation together as control measures (which was the situation when using Oman data) plays an important role to contain the disease. However, the isolation rate must be increased, especially institutionally, if quarantine rate is needed to be reduced for social and economic considerations.
\newpage
\bibliographystyle{plain}

\bibliography{BasicBib}
\end{document}